\documentclass[aps,twocolumn,floatfix,showpacs]{revtex4} %

\usepackage{amsmath,amsfonts,amssymb}
\usepackage{bbm}
\usepackage{color}
\usepackage{graphicx}

\usepackage{hyperref}

\usepackage{psfrag}
\usepackage{times}

\newsavebox{\inneroscillatorbox}
\newsavebox{\outeroscillatorbox}
\newsavebox{\surfaceoscillatorbox}
\newsavebox{\greenoscillatorbox}
\newsavebox{\blackoscillatorbox}
\newlength{\oscillatoroffset}
\newcommand{\oscillatorscale}{.45}
\savebox{\inneroscillatorbox}{\raisebox{\oscillatoroffset}%
{\includegraphics[scale=\oscillatorscale]{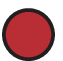}}}
\savebox{\outeroscillatorbox}{\raisebox{\oscillatoroffset}%
{\includegraphics[scale=\oscillatorscale]{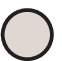}}}
\savebox{\surfaceoscillatorbox}{\raisebox{\oscillatoroffset}%
{\includegraphics[scale=\oscillatorscale]{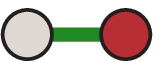}}}
\savebox{\greenoscillatorbox}{\raisebox{\oscillatoroffset}%
{\includegraphics[scale=\oscillatorscale]{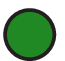}}}
\savebox{\blackoscillatorbox}{\raisebox{\oscillatoroffset}%
{\includegraphics[scale=\oscillatorscale]{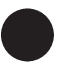}}}

\newcommand{\rr}{{\mathbbm{R}}}
\newcommand{\cc}{{\mathbbm{C}}}
\newcommand{\zz}{{\mathbbm{Z}}}

\newcommand{\id}{{\mathbbm{1}}}
\renewcommand{\vec}[1]{\boldsymbol{#1}}
\newcommand{\me}{\mathrm{e}}
\newcommand{\mi}{\mathrm{i}}

\newcommand{\arxiv}[1]{\href{http://arxiv.org/abs/#1}{#1}}
\newcommand{\coloneqq}{\ensuremath{\;\raisebox{.06ex}{:}\!\!=}}
\newcommand{\eqqcolon}{\ensuremath{=\!\!\raisebox{.06ex}{:}\;}}

\DeclareMathOperator{\circulant}{circ}
\DeclareMathOperator{\tr}{tr}

\begin{document}

\title{An entanglement-area law
for general bosonic harmonic lattice systems}

\author{M.~Cramer$^{1}$, J.~Eisert$^{2,3,1}$,  M.~B.~Plenio$^{2,3}$, and
J.~Drei{\ss}ig$^{1}$}

\affiliation{1 Institut f{\"u}r Physik, Universit{\"a}t Potsdam,
Am Neuen Palais 10, D-14469 Potsdam, Germany\\
2 QOLS, Blackett Laboratory, Imperial College London,
Prince Consort Road, London SW7 2BW, UK\\
3 Institute for Mathematical Sciences,
Imperial College London, Exhibition Road, London, SW7 2BW, UK}

\begin{abstract}

We demonstrate that the entropy of entanglement and the
distillable entanglement of regions with respect to the rest of a
general harmonic lattice system in the ground or a thermal
state scale at most as the boundary area of the region. This area
law is rigorously proven to hold true in non-critical harmonic lattice system
of arbitrary spatial dimension, for general finite-ranged harmonic
interactions, regions of arbitrary shape and states of nonzero
temperature. For nearest-neighbor interactions -- corresponding to
the Klein-Gordon case -- upper and lower bounds to the degree of
entanglement can be stated explicitly for arbitrarily shaped
regions, generalizing the findings of [Phys.\ Rev.\ Lett.~{\bf
94}, 060503 (2005)]. These higher dimensional
analogues of the analysis of block entropies in the
one-dimensional case show that under general conditions, one can
expect an area law for the entanglement in non-critical harmonic
many-body systems. The proofs make use of methods from
entanglement theory, as well as of results on matrix functions of
block banded matrices. Disordered systems are also considered. We
moreover construct a class of examples for which the two-point
correlation length diverges, yet still an area law can be proven
to hold. We finally consider the scaling of classical correlations
in a classical harmonic system and relate it to a quantum lattice
system with a modified interaction. We briefly comment on a
general relationship between criticality and area laws for the
entropy of entanglement.
\end{abstract}


\maketitle

\date{\today}
\section{Introduction}

Ground states of quantum systems with many constituents are
typically entangled.
In a similar manner as one can identify characteristic length
scales of correlation functions, quantum correlations are expected
to exhibit some general scaling behavior, beyond the details of a
fine-grained description. Such characteristic features provide a
physical picture that goes beyond the specifics of the underlying
microscopic model. A central question of this type is the
following: If one distinguishes a certain collection of
subsystems, representing some spatial
region, of a quantum many-body
system in a pure ground state, the state of this part will
typically have a positive entropy, reflecting the entanglement
between this region and the rest of the system \cite{Summers W
85,Stelmachovic B 01,Osterloh,NielsenPhase,Audenaert EPW
02,Latorre1,Korepin1,Fannes}. This degree of entanglement is certainly
expected to depend on the size and
 also on the shape of the region. Yet, how does the
degree of entanglement specifically depend on the size of the
distinguished region? In particular, does it scale as the volume
of the interior -- which is meant to be the number of degrees of
freedom of the interior?
Or, potentially as the area of the boundary, i.e., the number of
contact points between the interior and the exterior?

This paper provides a detailed answer to the scaling behavior of
the entanglement of regions with their exterior in a general
setting of harmonic bosonic lattice systems and provides a
comprehensive treatment of upper bounds on these quantities. We
find that in arbitrary spatial dimensions the degree of
entanglement in terms of the von~Neumann entropy scales
asymptotically as the area of the boundary of the distinguished
region. This paper significantly extends the findings of Ref.\
\cite{AreaPRL} on harmonic bosonic lattice systems: There, the
area-dependence of the geometric entropy has been  proven for
cubic regions in non-critical harmonic lattice systems of
arbitrary dimension with nearest-neighbor interactions,
corresponding to discrete versions of Klein-Gordon fields. In this
paper we extend our analysis to
 a general class of finite-ranged harmonic interactions
and also take regions of
arbitrary shape into account.
For thermal Gibbs states, the entropy of a reduction is longer a meaningful
measure of entanglement. Instead,
an area-dependence for an appropriate mixed-state entanglement measure,
the distillable entanglement, is established.
Also, an analogous statement holds for classical correlations in classical systems.
The area-dependence is even found in certain cases where
one can prove the divergence of the two-point correlation length.
This demonstrates that this previously conjectured dependence
between area and entanglement is valid under surprisingly general
conditions.

The presented analysis will make use of methods from the
quantitative theory of entanglement in the context of quantum
information science \cite{Plenio V 98,Eisert P 03,Plenio V 05}. It
has been become clear recently that on questions about scaling of
entropies and degrees of entanglement -- albeit often posed some
time ago -- new light can be shed with such methods \cite{Summers
W 85,Stelmachovic B 01,Osterloh,NielsenPhase,Audenaert EPW
02,Latorre1,Korepin1,Fannes,Keating,Cardy,GraphStates,Pachos P
04,Stabil,DMRG,Wolf VC 04,Botero R
04,Peschel,Beenakker,AreaPRL,Longrange,Wolf,Klich,Orus,Fine}. In
this language, quantum correlations are sharply grasped in terms
of rates that can be achieved in local physical transformations.
To assess quantum correlations using novel powerful tools from
quantum information and to relate them to information-theoretical
quantities constitutes an exciting perspective.

In the context of quantum field theory, such questions of scaling
of entropies and entanglement have a long tradition under the
keyword of geometric entropy. In particular, work on the geometric
entropy of free Klein-Gordon fields was driven in part by the
intriguing suggested connection \cite{Callan W 94} to the
Bekenstein-Hawking black hole entropy 
\cite{Bekenstein 72,Bekenstein 73,Bekenstein 04}. 
In seminal works by
Bombelli et al.~\cite{Bombelli KLS 86} and Srednicki
\cite{Srednicki 93} the relation between the entropy and
the boundary area of the region has been suggested and supplemented
with numerical arguments. This connection has been made more
specific using a number of different methods. In particular, for
half spaces in general and intervals in the one-dimensional case,
the problem has been assessed employing methods from conformal
field theory, notably
\cite{Holzhey LW 95,Preskill}, based on earlier work by Cardy and
Peschel \cite{CardyPeschel}, and by Cardy and Calabrese
\cite{Cardy}.

In one-dimensional non-critical chains, one observes a saturation
of the entanglement of a distinguished block, as was proven
analytically for harmonic chains \cite{AreaPRL,Audenaert EPW 02} and was
later observed for non-critical spin chains numerically
\cite{Latorre1} and analytically \cite{Korepin1,Keating,Cardy}.
In turn, in critical
systems systems, one often -- but not always -- finds a
logarithmically diverging entropy
\cite{Latorre1,Korepin1,Keating,Cardy}. In case of a model the
continuum limit of which leads to a conformal field theory, the
factor of the logarithmically diverging term is related to the
central charge of the conformal field theory. The findings of the
present paper and of the above-mentioned results motivate further
questions concerning the general area-dependence of the degree of
entanglement, for example in fermionic systems \cite{Wolf,Klich}.
In particular, the connection between the geometric entropy
fulfilling no area law and the correlated quantum many-body system
being critical is not fully understood yet. This is particularly
true for the interesting case of more than one-dimensional quantum
systems.

This paper is structured as follows. We start, in Section
\ref{Major}, with a presentation of the major results of the
paper. In Section~\ref{Prelims} we define our notation and recall
some basics on harmonic lattice systems. Section \ref{Bounds}
provides a general framework of upper and lower bounds for
entanglement measures, expressed in terms of the spectrum of the
Hamiltonian and two-point correlation functions. This analysis is
performed both for the case of the ground state as well as for
mixed Gibbs states. Of particular interest are Hamiltonians with
finite-ranged interactions in Section \ref{SectionFiniteRanged}.
For such Hamiltonians, we first study the behavior of the
two-point correlation functions, then the entanglement bounds. In
Section~\ref{Nomore}, discussing the Gibbs state case, we
determine temperatures above which there is no entanglement
left. A class of examples of Hamiltonians which exhibit a
divergent two-point correlation length in their ground state, but
an area-dependence of the entanglement is presented in
Section~\ref{Divergent} and expressed in analytical terms. The
specific case of Hamiltonians the interaction part of which can be
expressed as a square of a banded matrix is discussed in Section
\ref{ThaSquared}. In this case, very explicit expressions for
entanglement measures can be found. We then consider, in Section
\ref{Classical}, the case of classical correlations in classical
harmonic systems with arbitrary interaction structure.
Interestingly, this case is related to the quantum case for
squared interactions in the sense of Section~\ref{ThaSquared}.
Finally, we summarize what has been achieved in the present paper,
and present a number of open questions in this context.

\section{Main results}\label{Major}

Throughout the whole paper we will consider harmonic subsystems
on a $D$-dimensional cubic lattice $C=[1,\dots,n]^{\times D}$.
The system thus has $n^D$ canonical degrees of freedom.
The central question of this work will be: how does
the degree of entanglement of a distinguished region $I\subset C$
with the rest $O=C\setminus I$ scale with the size and shape of $I$?

\begin{figure}[tb]
\begin{center}
\includegraphics[scale=\oscillatorscale]{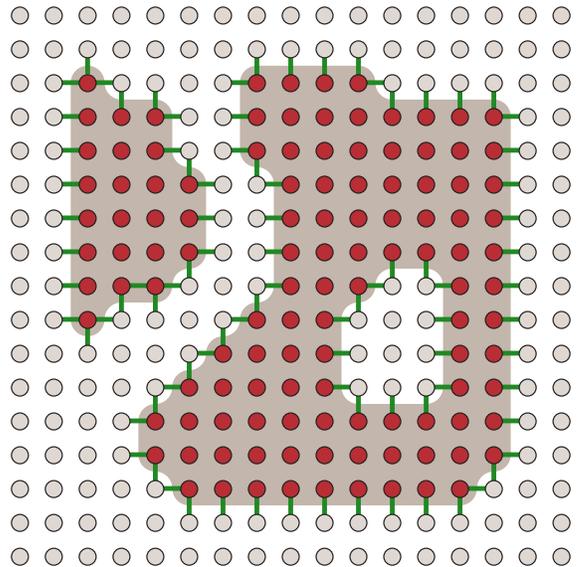}
\end{center}
\caption{A two-dimensional lattice system $C=[1,...,n]^{\times 2}$
with a distinguished (shaded) region $I$, consisting of $v(I)$
degrees of freedom~\usebox{\inneroscillatorbox}. Oscillators representing the
exterior $O=C\setminus I$ are shown
as~\usebox{\outeroscillatorbox},
whereas pairs belonging to the
surface $s(I)$ of the region are marked by lines (\usebox{\surfaceoscillatorbox}).
}\label{KnubbelSystem}
\end{figure}

We define the volume $v(I)$ and surface $s(I)$ of the
distinguished region $I$ as
\begin{equation*}
    v(I)=\sum_{\vec{i}\in I}1,\;\;\; s(I)=\sum_{\vec{i}\in
    O}
    \sum_{\substack{
           \vec{j}\in I \\
           d(\vec{i},\vec{j})=1}} 1,
\end{equation*}
where
\begin{equation*}
    d(\vec{i},\vec{j}) \coloneqq \sum_{\delta=1}^D| i_\delta-j_\delta|
\end{equation*}
defines the $1$-norm distance for vectors
$\vec{i}=(i_1,\dots,i_D)\in C$ that specify
the position of oscillators on the $D$-dimensional lattice.
More specifically, $s(I)$ is the number of contact points, that is, the
number of pairs of sites of $I$ and $O$ that are immediately adjacent. Note that the
distinguished region $I$ may have arbitrary shape and does not have to be
contiguous.

For the pure ground state $\varrho$, we will study the entropy of
entanglement of $I\subset C$ given by
\begin{equation*}
    E^S_{C,I} = S(\varrho_I),
\end{equation*}
where $S = -\tr[\rho\log \rho]$ is the von~Neumann entropy
of a state $\rho$ and $\varrho_I = \tr_{O}[\varrho]$ denotes the
reduced state associated with the degrees of freedom of the
interior $I$. For the pure ground state, this entropy of
entanglement is identical to both the distillable entanglement and
the entanglement cost. For pure states, it is indeed the unique
asymptotic measure of entanglement.
For Gibbs states, in turn,  we have to study a mixed-state
entanglement measure, as the entropy of a subsystem no longer
meaningfully quantifies the degree of entanglement. We will bound
the rate at which maximally entangled pairs can be distilled,
i.e., the distillable entanglement $E^{D}_{C,I}(T)$. Clearly,
$E^{D}_{C,I}(0)=E^{S}_{C,I} $ for zero temperature.

We will subsequently suppress the index $C$ for notational clarity. We
derive the following properties of $E^S_{I}$ and
$E^{D}_{I}(T)$: {\em
\begin{itemize}

\item[(I)] For $D$-dimensional harmonic lattice systems,
we derive general upper and lower bounds to $E_{I}^S$ for pure
ground states and to $E_{I}^D(T)$ for Gibbs states with respect to
a temperature $T>0$. These bounds are expressed entirely in terms
of the potential matrix and stated in Eqs.~(\ref{UpperBoundFiniteT}), (\ref{UpperBoundZeroT}), and
(\ref{LowerBound}).

\end{itemize}
}
\noindent A necessary condition for the following results to hold is that the spectral
condition number $\kappa=\lambda_\text{max}(V)/\lambda_\text{min}(V)$ of the coupling matrix $V$ satisfies $\kappa < c < \infty$ for some $c>0$ independent of $I$ and $O$.
{\em
\begin{itemize}
\item[(II)] For nearest-neighbor interactions and the ground state,
the entropy of entanglement $E^S_{I}$ scales
as the surface area $s(I)$
of $I$. More specifically,
there exist numbers $c_1,c_2>0$ independent of
$O$ and $I$ such that
\begin{equation*}
    c_1\cdot s(I) < E^S_{I}< c_2\cdot s(I).
\end{equation*}
This is stated in Eqs.~(\ref{FiniteUpperBound}) and
(\ref{NearestLowerBound}). Note that the specific case of cubic
regions $I=[1,\dots,m]^{\times D}$ has already been proven in the
shorter paper Ref.\
\cite{AreaPRL}.

\item[(III)] For general
finite-ranged harmonic interactions -- meaning arbitrary interactions which
are strictly zero after a finite distance -- the entropy of
entanglement $E^S_{I}$ of the ground state scales at most
linearly with the surface area of $I$: There exists a $c>0$
independent from $O$ and $I$ such that
\begin{equation*}
    E^S_{I}< c\cdot s(I).
\end{equation*}
This is expressed in Eq.~(\ref{FiniteUpperBound}).

\item[(IV)] For Gibbs states with respect to a temperature $T>0$
and general finite-ranged interactions the distillable
entanglement $E^D_{I}(T)$ scales at most linearly with the
surface area of $I$, so
\begin{equation*}
    E^D_{I}(T)< c(T)\cdot s(I),
\end{equation*}
with $c(T)>0$ being independent of $O$ and $I$, see again
Eq.~(\ref{FiniteUpperBound}), which applies for any temperature. We
also determine temperatures above which the entanglement is
strictly zero, see Eq.~(\ref{NoMoreTemp}).

\item[(V)] We construct a class of
Hamiltonians the ground state of which exhibit an
infinite two-point correlation length, for which the entropy of
entanglement $E^S_{I}$ scales provably at most linearly in the
boundary area. This is stated in Eq.~(\ref{DivergenceExample}).

\item[(VI)] For interactions that are specified by potential
matrices $V$ (see Eq.~(\ref{potmat})) that can be written as $V=W^2$
with $W$ corresponding to a finite-ranged interaction, we
determine the entropy of entanglement. This is made specific in
particular for $D=1$, see Eq.~(\ref{SquaredOneDim}).

\item [(VII)] For classical systems of harmonic oscillators
with nearest-neighbor interactions prepared in a thermal state,
the mutual information
\begin{equation*}
    I_{I}(T) = S_{I}(T) + S_{O}(T) - S_{C}(T)
\end{equation*}
measuring classical correlations scales linearly with the surface
area of the region. Here, $S_{I}$, $S_{O}$, and $S_{C}$ are
the discrete classical entropies in phase space of the interior,
the exterior, and the entire system with respect to an arbitrary
coarse graining as defined in Eq.~(\ref{Boltzmann}),
for a classical
Gibbs state at temperature $T>0$. The mathematical formulation of
the problem
is intimately related to case~(VI).\\
\end{itemize}
}

The remainder of the paper is now concerned with the detailed
discussion and rigorous proofs of these statements.

\section{Harmonic lattice systems -- Preliminaries}\label{Prelims}

We consider quantum systems on $D$-dimensional lattices, where
each site is associated with a physical system. The starting point
is the following Hamiltonian which is quadratic in the canonical
coordinates,
\begin{equation} \label{potmat} %
    H=\frac{1}{2}
    \Bigl(
    \sum_{\vec{i}} {p}_{\vec{i}}^2
    + \sum_{\vec{i},\vec{j}}{x}_{\vec{i}}V_{\vec{i},\vec{j}}{x}_{\vec{j}}
    \Bigr).
\end{equation}
The matrix $V$ is the potential matrix. $V_{\vec{i},\vec{i}}$ denotes
the potential energy contained in the degree of freedom labeled
with $\vec{i}$. For $\vec{i}\ne\vec{j}$ the element
$V_{\vec{i},\vec{j}}$ describes the coupling between the two
oscillators at $\vec{i}$ and $\vec{j}$. We assume that $V $ is a
real symmetric positive matrix. In this paper, we will consider
finite-ranged and nearest-neighbor interactions. Furthermore, interactions for which the correlation length diverges are considered.

We will subsequently study properties of both the ground state
$\varrho(0)$ as well as of Gibbs states
\begin{equation*}
        \varrho(T) = \frac{e^{- H/T}}{\tr [e^{-H/T}]}
\end{equation*}
corresponding to some nonzero temperature $T$. Note that we have
set the Boltzmann constant $k=1$. The questions we ask are
essentially those of the geometric entropy ($T=0$), and those on
the distillable entanglement ($T>0$) with respect to a
distinguished region $I$ of the lattice $C$. The specific case of
nearest-neighbor interactions, considered in Refs.\
\cite{Srednicki 93,Audenaert EPW 02,AreaPRL,Botero R 04}
corresponds to the one of a discrete version of free Klein-Gordon
fields in flat space-time. Here, we consider regions of arbitrary
shape and allow for more general interactions.

\subsection{Phase space, covariance matrices, and ground states}

The system we consider embodies $n^D$
canonical degrees of freedom,
associated with a phase space $(\rr^{\times2\,n^D},\sigma)$,
where the $(2\,n^D)\times (2\,n^D)$-matrix
$\sigma$,
\begin{equation*}
    \sigma \coloneqq
    \left[
    \begin{array}{cc}
    0_{n^D}  & \id_{n^D}\\
    -\id_{n^D} & 0_{n^D}
    \end{array}
    \right],
\end{equation*}
specifies the symplectic scalar product, reflecting the canonical
commutation relations between the $2\,n^D$ canonical coordinates
$\vec{o}=(x_1,\dots,x_{n^D}, p_1,\dots,p_{n^D})$ of position and
momentum. Instead of considering states we will refer to their
moments, considering that both the ground state as well as the
Gibbs states are quasi-free (Gaussian) states. The second moments
can be collected in the $(2\,n^D)\times (2\,n^D)$ real symmetric
covariance matrix $\gamma$ (see, e.g., Ref.~\cite{Eisert P 03}),
defined as
\begin{equation*}
        \gamma_{j,k} \coloneqq 2\Re (\tr[\varrho o_j o_k])=
    2 \Re \langle o_j o_k\rangle_{\varrho}.
\end{equation*}
Here, it is assumed that the first moments vanish, which does not
restrict generality as they can be made to vanish by local unitary
displacements that do not affect the entanglement properties.

Following Ref.~\cite{Audenaert EPW 02}, we find from symplectic
diagonalization that for the ground state at zero temperature
\begin{equation*}
        \gamma_0 = V^{-1/2} \oplus V^{1/2},
\end{equation*}
i.e., there is no mutual correlation between position and momentum,
and the two-point vacuum correlation functions are given by
the entries of $V^{-1/2}$
respectively $V^{-1/2}$,
\begin{align}
    G_{\vec{i},\vec{j}}&\coloneqq \langle
	\text{gs}|x_{\vec{i}}x_{\vec{j}}|\text{gs}\rangle=[V^{-1/2}]_{\vec{i},\vec{j}},
    \label{TwoPoint1}\\
    H_{\vec{i},\vec{j}}
    &\coloneqq \langle\text{gs}|p_{\vec{i}}p_{\vec{j}}|\text{gs}\rangle=[V^{1/2}]_{\vec{i},\vec{j}}.
\label{TwoPoint2}
\end{align}
Here, $|\text{gs}\rangle$ denotes the state vector of the ground state.
For the thermal Gibbs state at finite temperature the covariance matrix takes the form
\begin{equation*}
        \gamma(T) =  \left(V^{-1/2} W(T)\right) \oplus \left(V^{1/2} W(T)\right),
\end{equation*}
where we define
\begin{equation}\label{WT}
        W(T) \coloneqq   \id+ 2( \exp(V^{1/2}/T)- \id)^{-1}.
\end{equation}
Note that the additional term $W(T)$ is the same for the position
as well as for the momentum canonical coordinates. As in the zero
temperature case position and momentum are not mutually correlated
and $V^{-1/2} W(T)$ ($V^{1/2} W(T)$) are the two-point correlation
functions of position (momentum), now with respect to the Gibbs
state.

\subsection{Measures of entanglement}

The entropy of entanglement can be expressed in
terms of the symplectic eigenvalues of the covariance matrix corresponding to a reduction. Let the
$(2v(I)) \times (2v(I))$-matrix
$\left.\gamma_0\right|_I$ denote the covariance matrix associated with the interior $I$;
this is the principle submatrix of $\gamma_0$ associated with the degrees of freedom of the interior.
The symplectic spectrum of $\left.\gamma_0\right|_I$ is then defined as the spectrum
$\mu(B_I)$ of
the matrix
\begin{equation*}
    B_I\coloneqq  \left(\left.\gamma_0\right|_I\right)^{1/2} (\mi \left.\sigma\right|_I ) \left(\left.\gamma_0\right|_I\right)^{1/2}.
\end{equation*}
For the situation at hand we find that
\begin{equation*}
\left.\gamma_0\right|_I=\left.V^{-1/2}\right|_I\oplus \left.V^{1/2}\right|_I,
\end{equation*}
where we denote by $V^{\pm1/2}|_I$ the principle submatrix of $V^{\pm1/2}$ associated with the interior and we index the entries
of the submatrix as $[V^{-1/2}]_{\vec{i},\vec{j}}$ with vectors $\vec{i},\vec{j}\in I$.
Given the direct-sum structure of $\gamma_0|_I$ and counting doubly degenerate eigenvalues only once, we find that the entropy of entanglement can be evaluated as
\begin{equation}
    \label{entropy}
    E_{I}^S =\sum_{i=1}^{v(I)}
    \big[
    f\big(\frac{\mu_i- 1}{2}\big)
    -
    f\big(\frac{\mu_i+1}{2}\big)
    \big],
\end{equation}
where $f(x) = -x \log x $,
and the $\mu_i=\mu_i(A_I)$ are the square roots of the regular eigenvalues $\lambda_i=\lambda_i(A_I)$ of the $v(I)\times v(I)$ matrix $A_I$,
\begin{equation*}
\mu_i=\sqrt{\lambda_i(A_I)},\quad
A_I=\left.V^{-1/2}\right|_I\left.V^{1/2}\right|_I.
\end{equation*}
This reflects the fact that the entropy of a
harmonic system is the sum of the entropies of uncoupled degrees of freedom after symplectic diagonalization.\\
Using vectors $\vec{i},\vec{j}\in I$ to label the entries
of $A_I$, we find due to $V^{-1/2}V^{1/2}=\id$,
\begin{equation}\label{aMatrix}
\begin{split}
\left[A_I\right]_{\vec{i},\vec{j}}&=\left[\left.V^{-1/2}\right|_I\left.V^{1/2}\right|_I\right]_{\vec{i},\vec{j}}\\
&=\delta_{\vec{i},\vec{j}}-\sum_{\vec{k}\in O}\left[V^{-1/2}\right]_{\vec{i},\vec{k}}
    \left[V^{1/2}\right]_{\vec{k},\vec{j}}.
\end{split}
\end{equation}

\begin{figure}[tb]
\begin{center}
\psfrag{0}{$0$}
\psfrag{Ry30}[lb]{\hspace{0.2cm}$30$}
\psfrag{Ry35}[lb]{\hspace{0.2cm}$35$}
\psfrag{Ry65}[lb]{\hspace{0.2cm}$65$}
\psfrag{Ry70}[lb]{\hspace{0.2cm}$70$}
\psfrag{Rx30}{$30$}
\psfrag{Rx35}{$35$}
\psfrag{Rx65}{$65$}
\psfrag{Rx70}{$70$}
\psfrag{i}{$i$}
\psfrag{j}{$j$}
\includegraphics[width=\columnwidth]{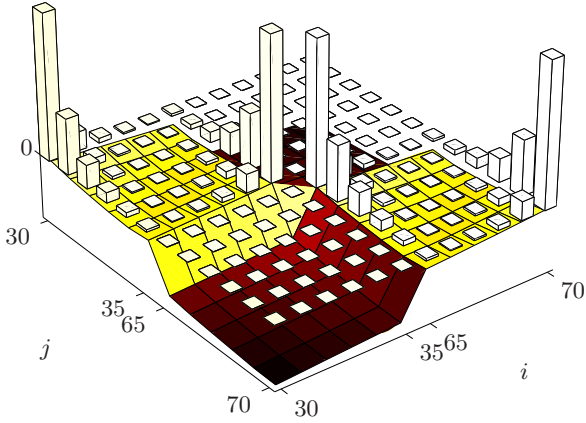}
\end{center}
\caption{Entries of $R$ in one dimension, $D=1$, 
for a finite-ranged coupling matrix $V$, $C=[1,...,100]$ and $I=[30,...,35]\cup[65,...,70]$,
yielding $s(I)=4$.
Bars show $R_{i,j}$, color-encoded surface 
depicts $\log(R_{i,j})$. All units are arbitrary. Note that the entries 
decay exponentially away from the boundary of $I$.}\label{R}
\end{figure}

This form of $A_I=\id-R$ hints at an area theorem in the following
way (see Fig.~\ref{R}, where we depict the entries of $R$): 
if the entries of $V^{\pm1/2}$ decay fast enough away from
the main diagonal, i.e., if the correlation functions decay
sufficiently fast, the main contribution to the entropy comes from
oscillators inside a layer around the surface of $I$ as for all
the others the product $[V^{-1/2}]_{\vec{i},\vec{k}}
[V^{1/2}]_{\vec{k},\vec{j}}$ will be very small (here vectors
$\vec{i},\vec{j}$ are inside $I$, $\vec{k}$ outside the region $I$).
Thus, the matrix $R$ has an effective rank proportional to $s(I)$, and 
the number of symplectic eigenvalues contributing to the sum in Eq.~(\ref{entropy}) 
is approximately proportional to $s(I)$.
Much of the remainder of the
paper aims at putting this intuition on rigorous grounds.

For mixed states, such as thermal Gibbs states, the von~Neumann entropy no longer represents a meaningful measure of the
present quantum correlations, and has to be replaced by concepts
such as the distillable entanglement. The distillable entanglement
is the rate at which one can asymptotically distill maximally
entangled pairs, using only local quantum operations assisted with
classical communication. For pure states, it coincides with the
entropy of entanglement -- the entropy of a reduction -- giving an
operational interpretation to this quantity \cite{Eisert P 03,
Plenio V 98,Plenio V 05}.

From entanglement theory we know that an upper bound for the
distillable entanglement is
provided by the logarithmic negativity \cite{Neg}. Note that this
is the case even in this infinite-dimensional context for Gaussian
states can be immediately verified on the level of second moments
and a single-mode description. Moreover, the entropy of
entanglement indeed still has the interpretation of a distillable
entanglement, albeit the fact that distillation protocols leave
the Gaussian setting. This is true as long as one includes an
appropriate constraint to the mean energy in the distillation
protocol \cite{InfContinuous}.

The logarithmic negativity \cite{Neg} is defined as
\begin{equation*}
        E_{I}^{N}(T) =
        \| \varrho(T)^\Gamma\|_1,
\end{equation*}
where $\|.\|_1$ denotes the trace norm, and $\varrho(T)^\Gamma$ is
the partial transpose of $\varrho(T)$ with respect to the split
$I$ and $O=C\setminus I$. Again following Ref.~\cite{Audenaert
EPW 02}, we find after a number of steps
\begin{equation}
\label{logneg}
        E^N_{I}(T)= \sum_{\vec{i}\in C}
\log_2\left[\max\left\{1,\lambda_{\vec{i}}\left(Q\right)\right\}\right],
\end{equation}
where $\lambda_{\vec{i}}(Q)$ labels the $n^D$ eigenvalues of $Q$,
\begin{equation*}
        Q\coloneqq P\omega^- P \omega^+,
\end{equation*}
and matrices $\omega^\pm(T)$ are defined as
\begin{equation*}
        \omega^\pm(T)\coloneqq W(T)^{-1}V^{\pm1/2},
\end{equation*}
which become $\omega^\pm=V^{\pm 1/2}$ for zero temperature
following Eq.~(\ref{WT}). The diagonal matrix $P$, defined as
\begin{eqnarray*}
P_{\vec{i},\vec{j}} &=& \delta_{\vec{i}\vec{j}}\left(\delta_{\vec{i}\in O}-\delta_{\vec{j}\in I}\right),\\
\delta_{\vec{i}\in S} &=&
 \begin{cases}
   1 & \text{if }\vec{i}\in S\\
   0  & \text{otherwise},
   \end{cases}
\end{eqnarray*}
is the matrix that implements time reversal in the subsystem
corresponding to the inner part $I$, reflecting partial
transposition $\varrho^\Gamma$ on the level of states.

\section{Upper and lower bounds}\label{Bounds}

In this section we will derive upper and lower bounds for the
entropy of entanglement and the distillable entanglement of the
distinguished region $I$ with respect to the rest of the lattice.
These bounds only depend on the geometry of the problem, i.e., the
region $I$ and on properties of $V$, namely its minimal and
maximal eigenvalue, which we define as $a\coloneqq \lambda_{\text{min}}(V)$
and $b\coloneqq\lambda_{\text{max}}(V)$, respectively,
its condition number
\begin{equation*}
    \kappa\coloneqq\frac{b}{a}=\frac{\lambda_{\text{max}}(V)}{\lambda_{\text{min}}(V)},
\end{equation*}
and the entries of $V^{\pm 1/2}$ respectively the
two-point correlation functions $G$ and $H$
as in Eqs.~(\ref{TwoPoint1},\ref{TwoPoint2}). In later sections these bounds
will be made specific for a wide range of interaction matrices
$V$.

\subsection{Upper bound}
\label{SectionUpperBound}

An upper bound for the entropy of entanglement and the distillable
entanglement is provided by the logarithmic negativity as in
Eq.~(\ref{logneg}). Utilizing this fact, we derive upper bounds using
$l_1$-norms \cite{Matrix} in this section. For brevity and
clarity, we will present the case for $T=0$ and finite temperature
in a single argument.

A direct calculation shows that
the matrix $Q$ is given by
\begin{eqnarray*}
        Q& =& \omega^- \omega^+ - 2 X \omega^+\\
    &=& W(T)^{-2}-2X\omega^+,
\end{eqnarray*}
introducing the matrix $X$ with entries
\begin{equation*}
        X_{\vec{i},\vec{j}} \coloneqq 
        \omega^-_{\vec{i},\vec{j}}
        \left(\delta_{\vec{i}\in I}
        \delta_{\vec{j}\in O}+\delta_{\vec{i}\in O}\delta_{\vec{j}\in
I}\right).
\end{equation*}
Therefore, we can bound the eigenvalues of $Q$ according to
\begin{eqnarray*}
        \lambda_{\vec{i}}(Q)
        &\le&
        \lambda_{\text{min}}(W(T))^{-2}
    +\lambda_{\vec{i}}(-2X\omega^+)\nonumber\\
        &\le&
        \lambda_{\text{min}}(W(T))^{-2}+2\left|
        \lambda_{\vec{i}}(X\omega^+)\right|,
\end{eqnarray*}
where we denote by $\lambda_{\text{min}}(W(T))$ the smallest
eigenvalue of $W(T)$ which is given by
\begin{equation*}
    \lambda_{\text{min}}(W(T))^{-1} =\lambda_{\text{max}}\Big(
    \frac{\me^{V^{1/2}/T}-\id}{\me^{V^{1/2}/T}+\id}\Big)
    =\frac{\me^{\sqrt{b}/T}-1}{\me^{\sqrt{b}/T}+1}.
\end{equation*}
Hence, we can write
\begin{eqnarray*}
        &&\hspace*{-0.75cm}E_{I}^N(T)
        \le\sum_{\vec{i}\in C}
        \log_2\left[\max\left\{1,\lambda_{\text{min}}(W(T))^{-2}+2\left|
        \lambda_{\vec{i}}(X\omega^+)\right|\right\}\right]
        \nonumber
        \\
        &\le&\frac{1}{\ln(2)}
        \sum_{\vec{i}\in C}
        \max\left\{0,\lambda_{\text{min}}(W(T))^{-2}-1+
        2\left|
        \lambda_{\vec{i}}(X\omega^+)\right|\right\},
\end{eqnarray*}
i.e., for $\lambda_{\text{min}}(W(T))^{-2}+2\max_{\vec{i}}|
\lambda_{\vec{i}}(X\omega^+)|<1$, there is no longer any
bi-partite entanglement in the system. We will later see that
there is a temperature $T_c$ above which this happens. But for
now, we use the fact that $\lambda_{\text{min}}(W(T))^{-2}
\le1$ to bound the
logarithmic negativity, and relate it to the $l_1$ norm
$\|.\|_{l_1}$ \cite{Matrix} of $ X $. We have
$\lambda_{\text{max}}(\omega_+)= \sqrt{b}/
\lambda_{\text{min}}(W(T))
$ and
therefore
\begin{multline*}
E_{I}^N(T)\le
        \frac{1}{\ln(2)}\sum_{\vec{i}\in C}2|
        \lambda_{\vec{i}}(X\omega^+)|=\frac{2}{\ln(2)}||X\omega^+||_1\\
    \le\frac{2\sqrt{b}}{\lambda_{\text{min}}(W(T))  \ln(2)}||X||_1 \le
    \frac{2\sqrt{b}}{\lambda_{\text{min}}(W(T)) \ln(2)}||X||_{l_1}.
\end{multline*}
The $l_1$ norm is defined as the sum of the absolute values of all
matrix entries. Inserting the definition of $X$, we find
\begin{equation}
\begin{split}
\label{UpperBoundFiniteT}
\frac{\lambda_{\text{min}}(W(T))
    \ln(2)}{2\sqrt{b}}E_I^D &\le ||X||_{l_1}=
        \sum_{\vec{i},\vec{j}\in C}
        \left|X_{\vec{i}, \vec{j}}\right|\\
    &=
       2 \sum_{\substack{
            \vec{j}\in I \\
            \vec{i}\in O}}
        \left|\omega^-_{\vec{i},\vec{j}}\right|
\end{split}
\end{equation}
for finite temperature, and
\begin{equation}
E_I^S
\le
\frac{4\sqrt{b}}{\ln(2)}
 \sum_{\substack{
    \vec{j}\in I \\
    \vec{i}\in O}}
        \left|[V^{-1/2}]_{\vec{i},\vec{j}}\right|
       \label{UpperBoundZeroT}
\end{equation}
for zero temperature.
These constitute upper bounds on the entropy of entanglement
and the distillable entanglement.
Both only depend in the distinguished region $I$,
the maximum eigenvalue of $V$, and the entries of $V^{-1/2}$, i.e.,
the two-point correlation function with entries
$G_{\vec{i},\vec{j}}$. This will be the starting point to
derive explicit upper bounds for special
types of interaction matrices $V$.

\subsection{Lower bound}

To achieve lower bounds for the zero temperature case, we consider
the entropy of entanglement directly. Starting from the general
expression Eq.~(\ref{entropy}), we use the fact that the
symplectic eigenvalues are never smaller than $1$, i.e., that the
eigenvalues of $A_I$ are contained in the interval $[1,\alpha_I]$,
with $\alpha_I$ being the maximal eigenvalue of $A_I$. Using the
pinching inequality \cite{Matrix}, we find
$\alpha_I\le(\lambda_{\text{max}}(V)/\lambda_{\text{min}}(V))^{1/2}
=(b/a)^{1/2}=\kappa^{1/2}$. Thus we can bound the entropy of
entanglement as follows,
\begin{eqnarray*}
    E_{I}^S  & \ge & \sum_{i=1}^{v(I)} \log [\mu_i]
    =\sum_{i=1}^{v(I)} \frac{\log [\lambda_i]}{2}
    \ge
    \frac{\log[\alpha_I]}{2(\alpha_I-1)}\sum_{i=1}^{v(I)}\left(\lambda_i-1\right)\\
    &\ge&
    \frac{\log[\sqrt{\kappa}]}{2(\sqrt{\kappa}-1)}\tr\left[A_I-\id\right].
\end{eqnarray*}
Using vectors $\vec{i},\vec{j}\in I$ to label the entries of the
$v(I)\times v(I)$ matrix $A_I$, we
finally arrive at the lower bound
\begin{eqnarray}
    E_{I}^S
    &\ge  & -\frac{\log[\sqrt{\kappa}]}{2(\sqrt{\kappa}-1)}
    \sum_{\substack{
      \vec{i}\in I \\
      \vec{j}\in O}}
    \left[V^{-1/2}\right]_{\vec{i},\vec{j}}
    \left[V^{1/2}\right]_{\vec{j},\vec{i}}.\;\;\;\;
    \label{LowerBound}
\end{eqnarray}
This lower bound depends only of the geometry of $I$, the spectral
condition number $\kappa$ of $V$ and the entries of $V^{\pm 1/2}$, i.e.,
the two-point correlation functions.

Eq.~(\ref{LowerBound}) is difficult to evaluate in general but
for special cases of the interaction matrix $V$ it can
nevertheless be made specific. In Section~\ref{NearestNeighbor},
for example, we give an explicit expression for the important case
of nearest-neighbor interactions. We expect Eq.~(\ref{LowerBound}) to be a convenient starting point to derive
such lower bounds also for more general cases of $V$.

A lower bound for the finite temperature case is difficult to obtain.
Generally speaking, a lower bound to the distillable entanglement
(two-way distillable entanglement in our case) is
given by the hashing inequality \cite{Hashing},
\begin{equation*}
    E_I^D(T) \geq
    \max\{
    S(\varrho_I ) - S(\varrho) ,
    S(\varrho_O ) - S(\varrho),0
    \},
\end{equation*}
where $\varrho_I = \tr_O[\varrho]$ and $\varrho_O= \tr_I
[\rho]$. Yet, naively applied, this inequality will vanish, as
generally $S(\varrho_I ) - S(\varrho)<0$ and $S(\varrho_O ) -
S(\varrho)<0$: this can be made intuitively clear from the
following argument. The above analysis demonstrates that both
the interior $I$ and the exterior $O$ can be approximately
disentangled with local unitaries up to a layer of the thickness
of the two-point correlation length. That is, degrees of freedom
associated with $\vec{i}\in O$
for which $d(\vec{i},\vec{j})$ is sufficiently
large for every $\vec{j}\in s(I)$,
can to a very good approximation be decoupled and
unitarily transformed into a thermal Gibbs state. Each
such degree of freedom will therefore contribute a
constant number to $S(\varrho)$, such that
$S(\varrho_I ) - S(\varrho)<0$ for sufficiently large $n$. Similarly,
one can argue to arrive at $S(\varrho_O ) -
S(\varrho)<0$. In order to establish a lower
bound to the distillable entanglement, however, we may start with
any protocol involving only local quantum operations and classical
communication, and apply the hashing inequality on the resulting
quantum states. This first step can include in particular local
filterings. A bound linear in the boundary area is expected to
become feasible if one first applies an appropriate unitary both
in $I$ and $O$, and then performs a local filtering involving
degrees of freedom associated with $\vec{i}\in I$ and $\vec{j}\in
O$ for which $d(\vec{i},\vec{j})=1$. This option will be explored
elsewhere.

\section{Finite-ranged interactions}\label{SectionFiniteRanged}

In this section we will make the upper bound on the entropy of
entanglement and the distillable entanglement explicit for
symmetric finite-ranged interaction matrices $V$, i.e., matrices
for which
\begin{equation*}
    V_{\vec{i},\vec{j}}=0,\;\text{for}\;d(\vec{i},\vec{j})>k/2,
\end{equation*}
where $d(\vec{i},\vec{j})$ denotes again the $1$-norm distance.
Denoting as before
the maximum and minimum eigenvalue of $V$ as
$a=\lambda_{\text{min}}(V)$ and $b=\lambda_{\text{max}}(V)$
respectively, we require that the spectral condition number
    $\kappa=b/a$
be strictly less than infinity independent of $C$, i.e.,
independent of $n$. Note that we do not require any further
assumptions on the matrix $V$.

\subsection{General upper bounds for finite-ranged interactions}

\begin{figure}[tb]
\begin{center}
\psfrag{0b}[cl]{\color{blue}$\scriptstyle 0$}
\psfrag{0}[cb]{$\scriptstyle 0$}
\psfrag{1}[cc]{\hspace{0.05cm}$\scriptstyle 0$}
\psfrag{4}[cc]{\hspace{0.05cm}$\scriptstyle 3$}
\psfrag{7}[cc]{\hspace{0.05cm}$\scriptstyle 6$}
\psfrag{10}[cc]{\hspace{0.1cm}$\scriptstyle 9$}

\psfrag{x3D}{$x_1$}
\psfrag{y3D}{$x_2$}

\psfrag{x}[c]{$\scriptstyle x_1$}

\psfrag{03D}{\hspace{0.2cm}$\scriptstyle 0$}
\psfrag{x1}{$\scriptstyle 0$}
\psfrag{x4}{$\scriptstyle 3$}
\psfrag{x7}{$\scriptstyle 6$}
\psfrag{x10}{$\scriptstyle 9$}

\psfrag{y1}{$\scriptstyle 0$}
\psfrag{y4}{$\scriptstyle 3$}
\psfrag{y7}{$\scriptstyle 6$}
\psfrag{y10}{$\scriptstyle 9$}
\includegraphics[width=\columnwidth]{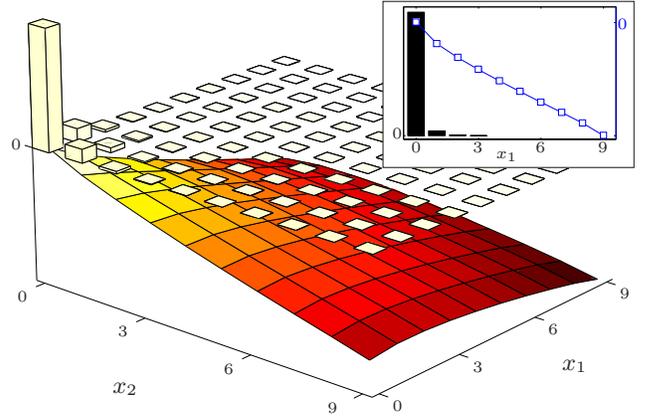}
\end{center}
\caption{The entries $[V^{-1/2}]_{\vec{i},\vec{j}}$ for a finite-ranged coupling
matrix $V$ in two dimensions, $D=2$, $\vec{j}=\vec{i}+(x_1,x_2)$. Bars show $[V^{-1/2}]_{\vec{i},\vec{j}}$
and the color-encoded surface shows $\log([V^{-1/2}]_{\vec{i},\vec{j}})$. The inset depicts the same for
$x_1=x_2$. All units are arbitrary. Note the exponential decay away from the main diagonal $\vec{i}=\vec{j}$.
}\label{sqrtinvV}
\end{figure}

We will make use of a result of Ref.~\cite{Benzi G 99} concerning
the exponential decay of entries of matrix functions. After generalizing this result to matrices
$V$ with the properties specified above it enables us to bound the
entries of $\omega^-$ as follows (see Appendix~\ref{decay})
\begin{equation}
\label{entryBounds}
        \left|\omega^-_{\vec{i},\vec{j}}\right|\le
        K_{a,b} \,
    q_{\kappa}^{d(\vec{i},\vec{j})},\;\;\;
        q_{\kappa}=\left(\frac{\kappa-1}{\kappa+1}\right)^{2/k},
\end{equation}
and
\begin{equation*}
    K_{a,b}=\frac{\me^{\eta/T}-1}{\me^{\eta/T}+1}
    \frac{\kappa+1}{\eta},
\end{equation*}
where
\begin{equation*}
    \eta \coloneqq  \left(a\frac{\kappa}{\kappa+1}\right)^{1/2}.
\end{equation*}
For zero temperature we then find
\begin{equation*}
    K_{a,b}=
    \frac{\kappa+1}{\sqrt{a}}\left({\frac{\kappa+1}{\kappa}}\right)^{1/2}.
\end{equation*}
This shows that off-diagonal terms of $V^{-1/2}$ decay
exponentially, cf.\ Fig.~\ref{sqrtinvV}.
Substituting Eqs.~(\ref{entryBounds})
into the general result in (\ref{UpperBoundFiniteT}), we find
\begin{equation*}
 \frac{\lambda_{\text{min}}(W(T))\ln(2)}{4 \sqrt{b}K_{a,b}} E_I^N(T)\le
        \sum_{\substack{
           \vec{i}\in I \\
           \vec{j} \in O}}
        q_{\kappa}^{d(\vec{i},\vec{j})}
        =
        \sum_{l=1}^\infty
        q_{\kappa}^l N_l,
\end{equation*}
where
\begin{equation*}
        N_l=\sum_{\vec{j} \in O}
        \sum_{\substack{
           \vec{i}\in I\\
           d(\vec{i},\vec{j})=l}}
        1.
\end{equation*}
Note that $N_1=s(I)$ coincides with the definition of the surface
area of $I$. We find $N_l\le 2(2l)^{2D-1}s(I)$ (see
Appendix~\ref{abzaehlerei}), i.e., we have
\begin{equation*}
        \sum_{\substack{
           \vec{i}\in I\\
           \vec{j} \in O}}
        q_{\kappa}^{d(\vec{i},\vec{j})}
        \le
        2^{2D}s(I)\sum_{l=1}^\infty q_{\kappa}^ll^{2D-1} \eqqcolon K_{D,\kappa}s(I).
\end{equation*}
Thus we finally arrive at the desired upper
bound linear in the surface area of $I$ for both the entropy of entanglement and the distillable entanglement,
\begin{equation}\label{FiniteUpperBound}
        E_{I}^N(T)\le
    \frac{4\sqrt{b}K_{a,b}K_{D,\kappa}}{\lambda_{\text{min}}(W(T))\ln(2)}s(I),
\end{equation}
which becomes
\begin{equation*}
        E_{I}^N\le
\frac{4(\kappa+1)^{\frac{3}{2}}K_{D,\kappa}}{\ln(2)}s(I)
\end{equation*}
for zero temperature, i.e., these upper bounds depend solely
on the maximal and minimal eigenvalue of the interaction matrix
$V$ and the surface area of the region $I$. This result
demonstrates that indeed, an area-bound of the degree of
entanglement holds in generality for bosonic harmonic lattice
systems. This shows that the previously expressed intuition can
indeed be made rigorous in form of an analytical argument.

\subsection{Disordered systems}

Notably, the derived results hold also for systems in which
the coupling coefficients are not identical, but
independent
realizations of random variables. If the coefficients
$V_{{\vec{i}},{\vec{j}}}$ of the real symmetric matrix $V$
are taken from a distribution with a carrier $[x_0,x_1]$,
such that the interaction is (i) finite-ranged, and (ii)
the carrier is chosen such that
\begin{equation*}
    \kappa = \frac{\lambda_{\text{max}}(V) }{
    \lambda_{\text{min}}(V) }<\infty,
\end{equation*}
then the same result holds true. This follows immediately from the
considerations in Appendix~\ref{decay}, where it is not assumed
that the Hamiltonian exhibits translational symmetry.
Eq.~(\ref{FiniteUpperBound}) is thus valid also for disordered harmonic
lattice systems.

\section{Temperatures above which there is no more entanglement
left}\label{Nomore}

In Section~\ref{SectionUpperBound} we found that for
$\lambda_{\text{min}}(W(T))^{-2}+2\max_{\vec{i}}|\lambda_{\vec{i}}(X\omega^+)|<1$
there is no entanglement between the regions $I$ and $O$. For
finite-range interactions $V$ we found in Section
\ref{SectionFiniteRanged}
\begin{eqnarray*}
\max_{\vec{i}}|\lambda_{\vec{i}}(X\omega^+)|
&\le& \lambda_{\text{min}}(W(T))^{-1}\sqrt{b}||X||_{l_1} \\
&\le& 2\lambda_{\text{min}}(W(T))^{-1} \sqrt{b}K_{a,b}K_{D,\kappa}s(I),
\end{eqnarray*}
 i.e., we have
\begin{multline*}
     \lambda_{\text{min}}(W(T))^{-2}+2\max_{\vec{i}}|\lambda_{\vec{i}}(X\omega^+)|
    \le\\
     \lambda_{\text{min}}(W(T))^{-2}+
    4\lambda_{\text{min}}(W(T))^{-1}
    \sqrt{b}K_{a,b}K_{D,\kappa}s(I).
\end{multline*}
On the right-hand side only $K_{a,b}$ and $\lambda_{\text{min}}(W(T))^{-2}$
depend on the temperature, both are decreasing in $T$ and go to zero as
$T$ goes to infinity, i.e.,
\begin{equation}\label{NoMoreTemp}
    \lambda_{\text{min}}(W(T))^{-2}+
    \frac{4\sqrt{b}K_{a,b}K_{D,\kappa}s(I)}{\lambda_{\text{min}}(W(T))}=1
\end{equation}
gives an implicit equation for the temperature $T_c$ above which
there is no bi-partite entanglement left in the system.

\section{Nearest-neighbor interactions}\label{Nearest}

\label{NearestNeighbor} In this section we consider
nearest-neighbor interactions and periodic boundary conditions in
$D$ spatial dimensions, i.e., block-circulant matrices $V$. For
$n\rightarrow\infty$ this is a special case of the matrices
considered in Section~\ref{SectionFiniteRanged} as boundary
conditions become irrelevant in this limit, i.e., the upper bound
coincides with the one derived for finite-ranged interactions. For
a tighter bound see Appendix~\ref{nnEntries}. We will now make use
of the circulant structure of $V$ to show that for matrices of
this kind it is possible to also derive a lower bound on the
entropy of entanglement  that it is proportional to the surface
area of $I$. We write $M=\circulant(\bar{M})$ for the circulant
matrix $M$ whose first block column is specified by the tupel of
matrices $\bar{M}$. We can then recursively define $V=V_D$ via
\begin{equation*}
    V_{\delta+1}=\circulant(V_{\delta},-c\id_{n^{\delta}},0_{n^{\delta}},\dots,0_{n^{\delta}},-c\id_{n^{\delta}}),
\end{equation*}
where $V_1=\circulant(1,-c,0,\dots,0,-c)$. We choose an energy
scale in which $V_{\vec{i},\vec{i}}=1$ and we demand $c>0$
\cite{plusminus}. This circulant structure leads to the following
properties of $V$: the eigenvalues of $V$ are given by
\begin{equation*}
\lambda_{\vec{i}}(V)=1-2c\sum_{\delta=1}^D\cos\left(2\pi i_\delta/n\right),\;\vec{i}\in C,
\end{equation*}
in particular the maximum eigenvalue is given by
$\lambda_{\text{max}}(V)\le 1+2cD$, where equality holds for $n$ even.
The minimum
eigenvalue reads $\lambda_{\text{min}}(V)=1-2cD$, i.e., positivity
of $V$ demands $c<1/2D$. 
Note that the assumption that $c<1/2D$ independent
of $I$ and $O$ is essential for the argument.
If we allow for an $n$-dependence of
$c$ as it arises, e.g., in the field-limit where $c=1/(1/n^2+2D)$,
then in one spatial dimension one will encounter an area law up to a
logarithmically divergent correction. This
behavior -- resembling the behavior of critical
spin chains and quadratic fermionic models --
will be studied in more detail elsewhere.

Furthermore the circulant nature of $V$
yields the following explicit expressions for the entries of
$V^{\pm 1/2}$
\begin{equation}
\label{entries}
\left[V^{\pm 1/2}\right]_{\vec{i},\vec{j}}=
\frac{1}{n^D}\sum_{\vec{k}\in C}\me^{2\pi\mi\vec{k}(\vec{i}-\vec{j})/n}\left(\lambda_{\vec{k}}(V)\right)^{\pm 1/2}.
\end{equation}
This reasoning leads to the following properties for
$\vec{i}\ne\vec{j}$ that are crucial for the present proof (see
Appendix~\ref{nnEntries}),
\begin{multline*}
-[V^{-1/2}]_{\vec{i},\vec{j}}[V^{1/2}]_{\vec{j},\vec{i}}
=\left|[V^{-1/2}]_{\vec{i},\vec{j}}[V^{1/2}]_{\vec{j},\vec{i}}\right|\\
\ge
\left(\frac{c}{2}\right)^{2d(\vec{i},\vec{j})}
\Big(\frac{d(\vec{i},\vec{j})!}{(2d(\vec{i},\vec{j})-1)\prod_{\delta=1}^D|i_\delta-j_\delta|!}\Big)^2.
\end{multline*}
Substituting these results into the general lower bound (\ref{LowerBound}) and keeping only terms with
$d(\vec{i},\vec{j})=1$ immediately yields
\begin{equation}\label{NearestLowerBound}
E^S_I\ge\frac{\log[\sqrt{\kappa}]c^2}{8(\sqrt{\kappa}-1)}s(I) ,
\end{equation}
where
\begin{equation*}
\kappa=\frac{1+2cD}{1-2cD}
\end{equation*}
in this case. This generalizes the result of Ref.\
\cite{AreaPRL} to regions $I$ of
 arbitrary shape.
We expect that these lower bounds can be generalized
to other interactions $V$ and numerical results suggest that these bounds hold quite
generally.

\section{Squared interactions}\label{ThaSquared}

A simple special case is related to a certain kind of
interaction: this the one where the interaction matrix $V$ can be written as
\begin{equation*}
    V=M^2
\end{equation*}
with a real symmetric matrix $M$ corresponding to a finite-range
interaction. Then, the covariance matrix associated with the
interior is nothing but
\begin{equation}\label{ga}
    \left.\gamma_0\right|_I= \left.M^{-1}\right|_I\oplus \left.M\right|_I.
\end{equation}
In this case, the symplectic spectrum can be determined in a fairly straightforward manner.
We mention this case also as it appears to
be an appropriate toy model as a starting
point for studies aiming at assessing the
symplectic spectrum of the reduction
itself and therefore the spectrum of the reduced density matrix \cite{Orus}.

We are looking for the spectrum of the $v(I)\times v(I)$ matrix $A_I$  as in
Eq.~(\ref{aMatrix}) which now takes the
form (recall that we label the entries of $A_I$ by vectors $\vec{i}, \vec{j}\in I$)
\begin{equation*}
\left[A_I\right]_{\vec{i},\vec{j}}=\delta_{\vec{i},\vec{j}}-\sum_{\vec{k}\in O}[M^{-1}]_{\vec{i},\vec{k}}
    M_{\vec{k},\vec{j}}
    =\delta_{\vec{i},\vec{j}}-R_{\vec{i},\vec{j}}.
\end{equation*}
It is now the central observation that the number of rows of the
matrix $R$ that are nonzero -- and therefore also the number of
eigenvalues that are nonzero -- is proportional to the surface
area of $I$, as $M$ is a banded matrix.
That is,
\begin{equation*}
    M_{\vec{i},\vec{j}}=0
\end{equation*}
for $d(\vec{i},\vec{j})>k/2$, meaning that
$R_{\vec{i},\vec{j}}=0$ if $d(\vec{k},\vec{j})>k/2$ for all $\vec{k}\in O$.
As the eigenvalues of $A_I$ are bounded through the pinching
inequality by $1$ and $\lambda_{\text{max}}(W)/\lambda_{\text{min}}(W)$,
an upper bound linear in the surface area of the interior follows immediately from the fact that the number of eigenvalues
entering the sum in Eq.~(\ref{entropy}) is proportional to $s(I)$. This argument
demonstrates that in this simple case, one immediately arrives at
bounds that are linear in the number of contact points.

\begin{figure}[tb]
\begin{center}
\psfrag{s0b}[cl]{\color{blue}$\scriptstyle 0$}
\psfrag{s0}[bc]{$\scriptstyle 0$}
\psfrag{0}{$\scriptstyle 0$}
\psfrag{Rsx1}{$\scriptstyle 1$}
\psfrag{Rsx4}{$\scriptstyle 4$}
\psfrag{Rsx7}{$\scriptstyle 7$}
\psfrag{Rsxa}{$\scriptstyle 11$}
\psfrag{Rsxb}{$\scriptstyle 14$}
\psfrag{Rsy1}{\hspace{0.2cm}$\scriptstyle 1$}
\psfrag{Rsy4}{\hspace{0.2cm}$\scriptstyle 4$}
\psfrag{Rsy7}{\hspace{0.2cm}$\scriptstyle 7$}
\psfrag{Rsya}{\hspace{0.2cm}$\scriptstyle 11$}
\psfrag{Rsyb}{\hspace{0.2cm}$\scriptstyle 14$}

\psfrag{q1}[lt]{$\scriptstyle 1$}
\psfrag{q4}[lt]{$\scriptstyle 4$}
\psfrag{q7}[lt]{$\scriptstyle 7$}
\psfrag{ qa}[lt]{$\scriptstyle 11$}
\psfrag{ qb}[lt]{$\scriptstyle 14$}
\psfrag{i}[cl]{$i$}
\psfrag{j}{$j$}
\includegraphics[width=\columnwidth]{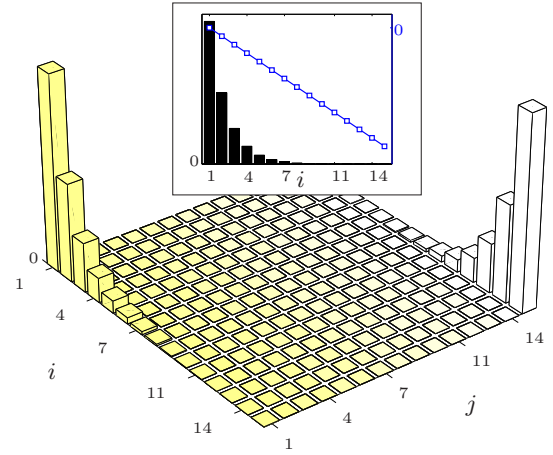}
\end{center}
\caption{Matrix entries of $R$ for the case $D=1$, $C=[1,...,100]$, $I=[1,...,15]$, $V=M^2$, where $M$ is a nearest-neighbor circulant matrix. Note that for this particular form of $V$ the entries $R_{i,j}$ are
exactly zero for $j\ne 1,m$. This is in contrast to the general form of
$R$ depicted in Fig.~\ref{R}. The inset shows the entries $R_{i,1}$
as bars and $\log(R_{i,1})$ in blue (cf.\ Eq.~(\ref{Rdecay})).}
\label{sq_R}
\end{figure}

The task of finding the eigenvalues of $A_I$ explicitly is now reduced to finding the
eigenvalues of the sparse matrix $R$,
\begin{equation*}
\mu_{\vec{i}}(A_I)=\sqrt{1-\lambda_{\vec{i}}(R)}.
\end{equation*}
This case is particularly transparent in the one-dimensional case,
$D=1$, and for $M$ being a circulant matrix with first row $(1,
-c,0,\dots,-c)$ for $0<c<1/2$. The potential matrix $V$
corresponds then to nearest-neighbor interactions, together with
next-to-nearest-neighbor interactions. In this one-dimensional
setting, we set $I=[1,\dots,m]$. The matrix $R$ then takes the
simple form (cf.\ Fig.~\ref{sq_R})
\begin{eqnarray*}
    R_{i,j}&=&\sum_{k=m+1}^n[M^{-1}]_{i,k}M_{k,j}\\
    &=&-c\left([M^{-1}]_{i,m+1}\delta_{j,m}+[M^{-1}]_{i,n}\delta_{j,1}\right).
\end{eqnarray*}
To find its eigenvalues, we calculate $\det(R-\lambda\id)$, which is
straightforward for a matrix of this form.
\begin{eqnarray*}
    \det(R-\lambda\id)=-(-\lambda)^{m-2}\det
\left( 
\begin{array}{cc}
R_{11}-\lambda & R_{1m}\\
R_{m1} & R_{mm}-\lambda
\end{array}
\right)\\
=(-\lambda)^{m-2}(R_{1m}R_{m1}-
(R_{11}-\lambda)(R_{mm}-\lambda)),
\end{eqnarray*}
reflecting the fact that $m-2$ eigenvalues are zero, i.e, the number
of nonzero eigenvalues is $s(I)=s([1,\dots,m])=2$. We find for the non-vanishing eigenvalues
\begin{eqnarray*}
    2\lambda_\pm &=&R_{11}+R_{mm}\\
    &\pm&
    \left(
    (R_{11}-R_{mm})^2
    +4R_{1m}R_{m1}\right)^{1/2}.
\end{eqnarray*}
Symmetry of $M^{-1}$ yields
\begin{equation*}
\lambda_\pm=R_{11}\pm R_{1m}
=-c\left([M^{-1}]_{1,n}\pm [M^{-1}]_{m,n}\right).
\end{equation*}
From the circulant structure of $M$ we have
\begin{eqnarray*}
[M^{-1}]_{i,j}&=&
\frac{1}{n}\sum_{k=1}^n\frac{\me^{2\pi\mi k(i-j)/n}}{1-2c\cos(2\pi k/n)}\\
&=&
\frac{1}{n}\sum_{k=1}^n\frac{\cos(2\pi k(i-j)/n)}{1-2c\cos(2\pi k/n)},\\
\end{eqnarray*}
i.e., for large $n$ we arrive at
\begin{eqnarray}
[M^{-1}]_{1,n}&=&\frac{1}{2c\sqrt{1-4c^2}}-\frac{1}{2c},\nonumber\\\
[M^{-1}]_{m,n}&=&\frac{1}{\sqrt{1-4c^2}}\left(\frac{1-\sqrt{1-4c^2}}{2c}\right)^m.\label{Rdecay}
\end{eqnarray}
Note that these expressions are
asymptotically independent of $m$ as shown in the inset of Fig.~\ref{sq_R}. In this limit
we finally arrive at
\begin{equation}\label{SquaredOneDim}
\lambda_\pm=\frac{1}{2}-\frac{1}{2\sqrt{1-4c^2}}.
\end{equation}
This expression specifies the symplectic spectrum of the
reduction in a closed form.

\section{Diverging correlation length and area-law of the entropy}\label{Divergent}

In this section we will analytically demonstrate that
there exist Hamiltonians for which
the ground state two-point correlation functions diverge
whereas the geometric entropy
is still bounded by an area-law.
In spin systems with a long-range Ising interaction
such a behavior has been observed in the one-dimensional
case and sketched for higher dimensions \cite{Longrange}.
Here, we present a
class of examples of this type valid in arbitrary dimension
$D$, for which one can prove the validity of an upper bound
linear in the boundary area.
Moreover, this set of examples is not restricted
to cubic regions. The interaction is here a suitable
harmonic long-range interaction.
This analysis shows that a divergent two-point
correlation function alone is no criterion for a
saturating block entropy in the one-dimensional case, and
for an area-dependence in higher dimensions.

Consider the matrix $M$ with entries
\begin{eqnarray*}
    M_{\vec{i},\vec{j}}& =& \frac{1}{ d(\vec{i},\vec{j})^\alpha},\\
    M_{\vec{i},\vec{i}} & = & 1+\sum_{\substack{
     \vec{j}\in C\\
     \vec{j}\ne\vec{i}}}
    \frac{1}{d(\vec{j},\vec{i})^\alpha}
\end{eqnarray*}
for some $\alpha >0$.
Now set $V=M^{-2}$. This choice implies that
the correlation function
$G_{\vec{i},\vec{j}}=[V^{-1/2}]_{\vec{i},\vec{j}}$ decays only
algebraically. We will show that despite this fact one still has an
upper bound linear in the boundary area of $I$
for appropriate values of $\alpha$.

Firstly, we have to make sure that the maximum eigenvalue of $V$
can be bounded from above independent of $n$. From Gershgorin's
theorem (see, e.g., Ref.\ \cite{Matrix})
we know that for every eigenvalue
$\lambda(M)$ there exists an ${\vec i}$ such that
\begin{equation*}
    |\lambda(M)-M_{\vec{i},\vec{i}}|\le
    \sum_{\substack{\vec{j}\in C\\ \vec{j}\ne\vec{i}}}
    M_{\vec{i},\vec{j}},
\end{equation*}
i.e.,
\begin{equation*}
    \lambda_{\text{min}}(M)\ge \min_i \Big[
    M_{\vec{i},\vec{i}}-\sum_{\substack{
    \vec{j}\in C\\
    \vec{j}\ne\vec{i}}}
    M_{\vec{i},\vec{j}}\Big]=1,
\end{equation*}
and therefore
\begin{equation*}
b=\lambda_{\text{max}}(V)=
\lambda_{\text{max}}(M^{-2})=\frac{1}{\lambda_{\text{min}}^2(W)}\le 1.
\end{equation*}
Substituting this and the specific form of $M$ into the general expression
for the entropy, we obtain
\begin{eqnarray*}
\frac{\ln(2)}{4}E_I^S&\le&
\sum_{\substack{\vec{j}\in I\\ \vec{i}\in O}}\frac{1}{\left[d(\vec{i},\vec{j})\right]^\alpha}
=\sum_{l=1}^\infty \frac{1}{l^\alpha}N_l\\
&\le& 2^{2D}s(I)\sum_{l=1}^\infty l^{2D-\alpha-1},
\end{eqnarray*}
which converges for $\alpha>2D$ to
\begin{equation}\label{DivergenceExample}
    E_I^S\le
    \frac{2^{2D+2}}{\ln(2)}\zeta(\alpha-2D+1)s(I),
\end{equation}
where $\zeta(x)$ is the Riemann zeta function. Note that these
bounds are not necessarily tight in the sense that even for
smaller values of $\alpha$, such a behavior can be expected. Steps
towards tightening these bounds seem particularly feasible in case
of cubic regions, where we conjecture that for $\alpha>D$ we
arrive at an area-dependence. This analysis shows that for
long-range interactions, an area-law in the degree of entanglement
can be concomitant with divergent two-point correlation functions.

\section{An area law for classical correlations}\label{Classical}

In previous sections we have considered the entanglement between some region
and the rest of a lattice of interacting
quantum harmonic oscillators with
Hamiltonians that are at most quadratic in position and momentum.  We have demonstrated
that both for the ground state and the thermal state of the entire lattice the quantum correlations,
i.e., the amount of
entanglement, between the region and the remainder of the lattice is bounded by quantities
proportional to their boundary surface
area.

This suggests similar questions concerning the
classical correlations between a region and the rest of the lattice in the
corresponding classical systems when the lattice system as a whole is prepared
in a thermal state. In this section we will
demonstrate that indeed analogous area laws hold.
We will note furthermore that there is a quite striking intimate relation
between the {\it classical system} with potential matrix $V_c$ and the
{\it quantum system} with a potential matrix
$V_q=V_c^2$ in this context.

\subsection{Hamiltonian, entropy, and mutual information}

For the following considerations we use the classical equivalent of the quantum
mechanical Hamiltonian (\ref{potmat}), namely again
\begin{equation} \label{ClassicalHam}
    H=\frac{1}{2}
    \Bigl(
    \sum_{\vec{i}} {p}_{\vec{i}}^2
    + \sum_{\vec{i},\vec{j}}{x}_{\vec{i}}V_{\vec{i},\vec{j}}{x}_{\vec{j}}
    \Bigr),
\end{equation}
where now $x=(x_1,\ldots,x_{n^D})$ and $p=(p_1,\ldots,p_{n^D})$ are
vectors of classical position and momentum variables, respectively, and
$V$ denotes again the potential matrix. The state of the classical
system is characterized by a phase space density $\varrho= \varrho(\xi)$,
a classical probability distribution, where $\xi =(x_1,\ldots,x_{n^D},p_1,\ldots,p_{n^D})$
denotes all the canonical coordinates in phase space.
For nonzero temperatures $T>0$
this phase space distribution
is given by the Boltzmann distribution
\begin{equation}\label{density}
    \varrho({\xi}) = \frac{1}{Z} e^{-\beta H(\xi)}
\end{equation}
where $\beta=1/T$ (as before, $k=1$), and
\begin{equation*}
    Z \coloneqq  \int d\xi e^{-\beta H(\xi)}.
\end{equation*}

Given this density in phase space we will encounter a familiar
ambiguity when defining the entropy of the system. Using the
discrete classical entropy \cite{Wehrl}, we split the phase space
into cubic cells each with a volume $h^{2N}$, where $N=k^D$ and
$h>0$ being an arbitrary constant. From the phase space density we
obtain the probability associated with each of these cells which
in turn can be used to determine the entropy function of this
probability distribution. We will now make use of the multiple
indices ${\vec i}=(i_1,\ldots,i_{N})$ and ${\vec
j}=(j_1,\ldots,j_{N})$, assuming that the cell corresponding to
$({\vec i},{\vec j})$ is centered around $x=(h i_1,\ldots,h i_N)$
and $p=(h j_1,\ldots,h j_N)$. That is, for each degree of the $n^D$
degrees of freedom, the phase space is discretized. The
contribution of each cell to the discretized probability density
is then given by
\begin{equation*}
    p({\vec i},{\vec j}) = \int_{\text{Cell}}
    d\xi \varrho(\xi) \, .
\end{equation*}
As usual, the discrete classical entropy is then defined as
the corresponding Shannon entropy
\begin{equation}\label{Boltzmann}
    S_C(h) \coloneqq  -\sum_{{\vec i},{\vec j}}
     p ({\vec i},{\vec j}) \log p ({\vec i},{\vec j})
\end{equation}
We will denote the discrete classical entropy with respect to the degrees of freedom of the interior with $S_{I}(h)$, the entropy of the exterior with $S_{O}(h)$.

The value for the entropy will depend on the choice of $h$ and in the limit $h\rightarrow 0$ this entropy
definition will diverge due to a term proportional to $-\log (h^{2N}) $. In classical statistical mechanics this problem is
avoided with the help of the third law of thermodynamics.
The entropy itself is however not the quantity that we wish to compute but rather the mutual information
$I_{I}$ between the
interior $I$ and the rest of a lattice, denoted as before by $O$.
This classical mutual information meaningfully quantifies
the classical correlations between the inner and the outer. In that case we find that the limit $h\rightarrow 0$ exists and
that the mutual information $I$ can be defined as
\begin{equation*}
    I_{I} \coloneqq \lim_{h\rightarrow 0}
    \left(
    S_{I}(h)
    + S_{O}(h)
    - S_{C}(h)
    \right).
\end{equation*}
Following these preparations we are now in a position to determine the mutual information between a region and the rest of
the lattice explicitly when the lattice as a whole is in a thermal state.

\subsection{Evaluation of the mutual information}

For the evaluation of the mutual information we need to determine
the entropy of the total lattice $C$, as well as the entropy
determined by the reduced densities describing the two regions $I$
and $O$. To this end we carry out the partial summation over all
degrees of freedom of region $O$ in order to find the reduced
phase space density $\varrho_I$ describing region $I$ only. Employing
the Schur complement we find that the reduced density $\varrho_I$ is
described by the Boltzmann distribution corresponding to the same
temperature and the Hamiltonian
\begin{equation} \label{ClassicalI}
     H_I = \frac{1}{2}\left(
    \sum_{\vec{i,j}\in I}
    x_{\vec{i}}
    [(\left.V\right|_I)^{-1}]_{\vec{i},\vec{j}}
    x_{\vec{j}}
    +
    \sum_{\vec{i}\in I}
    p_{\vec{i}}^2\right)
\end{equation}
An analogous result
holds for the reduced phase space density of region $O$.

For a thermal phase space distribution Eq.~(\ref{density})
corresponding to a classical Hamiltonian function of the form
Eq.~(\ref{ClassicalHam}) we can compute the entropy straightforwardly,
to find
\begin{equation*}
    S_{A} = -\frac{1}{2} \log \det \left(\left.V\right|_A\right)^{-1}
    + v(A) \log \frac{2\pi}{\beta} + v(A)
\end{equation*}
$A\in\{I,O,C\}$, which increases with temperature as expected. For the mutual information we find
\begin{equation*}
    I_{I} = \frac{1}{2} \log \frac{\det \left.V\right|_I \det \left.V\right|_O}{\det V},
\end{equation*}
which is, perhaps surprisingly, independent of temperature. Using Jacobi's determinant identity
\begin{equation*}
    \det \left.V\right|_O\det V^{-1}=\det \left.V^{-1}\right|_I
\end{equation*}
 this expression can be rewritten as
\begin{equation*}
    I_{I}  = \frac{1}{2} \log
    \det\left[ \left.V\right|_I \left.V^{-1}\right|_I\right]=
    \frac{1}{2} \log
    \det\left[ \id -R\right].
\end{equation*}
It is now advantageous to notice the close connection of this expression, in particular of the matrix $R$
\begin{equation*}
    R_{\vec{i},\vec{j}}=\sum_{\vec{k}\in O}[V^{-1}]_{\vec{i},\vec{k}}V_{\vec{k},\vec{j}}
\end{equation*}
with those that arise in the quantum mechanical problem that we
have treated previously in Section~\ref{ThaSquared}. Indeed, the
classical problem for a system with potential matrix $V_\text{c}$ is
related to the quantum mechanical system with the squared
potential matrix $V_\text{q} = V_\text{c}^2$. This formal similarity arises
because in Section~\ref{ThaSquared} we have shown that for a
lattice of quantum harmonic oscillators with potential matrix
$V_\text{q}=V_\text{c}^2$ in its ground state the symplectic eigenvalues of the
covariance matrix describing region I alone are exactly the
standard eigenvalues of the matrix $(\id - R)^{1/2}$. The
properties of these eigenvalues have already been discussed in
detail in Section~\ref{ThaSquared}. This allows us now
straightforwardly to establish the area theorem for the mutual
information in a classical system employing the result for the
corresponding quantum system.

This establishes in particular that the classical correlations as
measured by the mutual information $I_{C,I}$ between the
distinguished region $I$ and the rest of the lattice
$O=C\setminus I$ satisfy
\begin{equation*}
        c_1'\cdot s(I)\le
        I_{I} (T)\le c_2'\cdot s(I)
\end{equation*}
for large $v(I)$ and appropriate
constants $c'_{1,2}>0$ independent of $O$ and $I$.
(Compare also the assessment of the
thermodynamical entropy of parts
of classical fluids in Ref.\
\cite{Callaway}.) In
summary we have seen that the area-dependence of
correlations is not restricted to quantum systems, as long as one 
replaces the notion of entanglement -- representing
quantum correlations -- by the notion of classical correlations in a classical system.

\section{Summary and outlook}
In this paper, we have considered the question of the
area-dependence of the geometric entropy and the distillable
entanglement in general bosonic harmonic lattice systems of
arbitrary dimension. The question was the general scaling behavior
of these measures of entanglement with the size of a distinguished
region of a lattice. Such an analysis generalizes assessments of
block entropies in the one-dimensional case. Using methods from
entanglement theory, we established bounds that allow for a
conclusion that may be expressed in a nutshell as: 
{\it in surprising generality, we find that the degree
of entanglement scales at most linearly in the boundary area of
the distinguished region.}
This analysis shows that the intuition
that both the interior and the exterior can be approximately
disentangled up to a layer of the thickness of the two-point
correlation length by appropriate local unitaries carries quite
far indeed. 

For cubic regions $I=[1,...,m]^{\times D}$ the area law can be formulated as
\begin{equation}\label{TheArea}
    E^D_I=\Theta(m^{D-1}),
\end{equation}
where $\Theta$ is the Landau theta.

Such area-laws are expected
to have an immediate implication on the accuracy to which ground
states can be approximated with matrix-product states and higher
dimensional analogues in classical simulations of the ground
states of quantum many-body systems \cite{DMRG}. After all, the
failure of DMRG algorithms close to critical points can be related
to the logarithmic divergence of the block entropy in the
one-dimensional case.

The findings of the present paper raise a number of interesting
questions. Notably, in general quantum many-body systems on a
lattice (fermionic or bosonic), what are necessary and sufficient conditions for an
area-law in the above sense to hold? Clearly, as we have seen
above, the divergence of two-point correlation functions alone is
not in one-to-one correspondence with an area law. It would be
interesting to consider and
possibly decide the conjecture that a one-to-one
relationship between a system being critical and not satisfying a
law of the form as in Eq.~(\ref{TheArea}) holds if one (i)
restricts attention to systems in arbitrary dimension with
nearest-neighbor interactions, and (ii) grasps criticality in
terms of two-point correlation functions with algebraic decay,
concomitant with a vanishing energy gap. Note that the latter two
criteria of criticality have not been rigorously related to each
other yet and may indeed not be simultaneously satisfied in
lattice systems. The general relationship is still awaiting
rigorous clarification.

As steps towards such an understanding of a relationship between
criticality and properties of ground state entanglement in more
than one-dimensional fermionic and bosonic systems, it seems very interesting to study
models different from the ones considered in this paper. For example,
complementing our bosonic analysis,
area-laws in fermionic critical systems have been addressed
\cite{Wolf,Klich}, where logarithmic corrections have been
found also in higher-dimensional settings. Other settings in the bosonic
case, corresponding to field theories beyond this quasi-free setting,
are also still not clarified. In particular, the
scaling behavior of the geometric entropy in general bosonic
theories in higher dimensions is far from clear.
Then, the case of finite-size
effects in harmonic lattice systems where the correlation length
is larger than the full system, resembling the critical
case,  will be presented elsewhere. In less generality, it seems
also feasible to identify the prefactors of the leading and next-to-leading order
terms in an area-law of the geometric entropy. It is the hope that
the present work can contribute to an understanding of
genuine quantum correlations in quantum
many-body systems and inspire such further considerations.

\section{Acknowledgments}

We would like to thank C.~H.~Bennett, H.~J.~Briegel, J.~I.~Cirac,
W.~D{\"u}r,
M.~Fleischhauer, 
B.~Reznik,
T.~Rudolph, N.~Schuch, 
R.~G.~Unanyan,
R.~F.~Werner, and M.~M.~Wolf for
discussions. This work has been supported by the EPSRC QIP-IRC
(GR/S82176/0), the EU Thematic Network QUPRODIS (IST-2002-38877),
the DFG (SPP 1078 and SPP 1116), The Leverhulme Trust (F/07 058/U)
and the European Research Councils (EURYI).

\appendix

\section{Exponential decay of entries of matrix functions}

\label{decay}

The result concerning the exponential decay of entries of matrix functions of Ref.\
\cite{Benzi G 99} relies on the fact that the $p$-th power $A^p$
of a $k$-banded matrix $A=(A_{i,j})$ is $pk$-banded, i.e., 
$[A^p]_{i,j}=0$ for $|i-j|>pk/2$ for a matrix $A$ with $A_{i,j}=0$ 
for $|i-j|>k/2$. For the purposes of the present paper, we will 
need a generalization of the result of Ref.\ \cite{Benzi G 99}
to block banded matrices. We refer to $V=(V_{\vec{i},\vec{j}})$
as being $k$-banded, if
$V_{\vec{i},\vec{j}}=0$ for $d(\vec{i},\vec{j})>k/2$. 
It can be proven
by induction over $p$ that the $p$-th power of $V$ is $pk$ banded
in this sense. 
This enables us to formulate the general form of Ref.\
\cite{Benzi G 99} as follows.

Let $V=(V_{\vec{i},\vec{j}})$ be a $k$-banded symmetric matrix, i.e., $V_{\vec{i},\vec{j}}=0$ for
$\sum_\delta|i_\delta-j_\delta|>k/2$.
Define $a=\lambda_{\text{min}}(V)$, $b=\lambda_{\text{max}}(V)$,
$\kappa=b/a$,
\begin{equation*}
    \psi : \cc\rightarrow\cc,\;\;\; \psi(z) =\frac{(b-a)z+a+b}{2},
\end{equation*}
and $\varepsilon_\chi$ as an ellipse with foci in
$-1$ and $1$ and half axes $\alpha$, $\beta$, $\chi=\alpha+\beta$.\\
Now let $f:\cc\rightarrow\cc$ be such that $f\circ \psi$ is
analytic in the interior of the ellipse $\varepsilon_\chi$, $\chi>1$,
and continuous on $\varepsilon_\chi$.
Furthermore suppose $(f\circ \psi)(z)\in \rr$ for $z \in \rr$.
Then there exist constants $K$ and $q$, $0\le K$, $0\le q\le 1$
such that
\begin{equation*}
\left|[f(V)]_{\vec{i},\vec{j}}\right|
\le
K q^{\sum_{\delta=1}^D|i_\delta-j_\delta|},
\end{equation*}
where
\begin{gather*}
K=\max\left\{\max_{\vec{i}}\left|\lambda_{\vec{i}}(f(V))\right|,\frac{2\chi
M(\chi)}{\chi-1}\right\},\label{K}\\
q=\left(\frac{1}{\chi}\right)^{2/k},\quad
M(\chi)=\max_{z\in\varepsilon_\chi}|( f\circ \psi)(z)|.\nonumber
\end{gather*}
To bound the entries of $\omega^-$, we apply the above theorem to the function
\begin{equation*}
\omega^-(z)=\frac{\me^{\sqrt{z}/T}-1}{\me^{\sqrt{z}/T}+1}z^{-\frac{1}{2}}.
\end{equation*}
For $1<\chi<(\sqrt{\kappa}+1)^2/(\kappa-1)$, we have that $\omega^- \circ \psi$ is analytic in the interior of the ellipse $\varepsilon_\chi$ and continuous on $\varepsilon_\chi$, i.e.,
$\omega^-$ satisfy the assumptions of the above theorem.
To make $K$ and $q$ specific,
we choose
\begin{equation*}
    \frac{(\sqrt{\kappa}+1)^2}{\kappa-1}>\chi\coloneqq \frac{\kappa+1}{\kappa-1}>1,
\end{equation*}
which yields
\begin{equation*}
    q_{\kappa}=\left(\frac{\kappa-1}{\kappa+1}\right)^{2/k},\quad
    K_{a,b}=\frac{\me^{\eta/T}-1}{\me^{\eta/T}+1}
    \frac{\kappa+1}{\eta}
\end{equation*}
where
\begin{equation*}
    \eta=\left(a\frac{\kappa}{\kappa+1}\right)^{1/2},
\end{equation*}
and for zero
temperature $K_{a,b}$ reduces to
\begin{equation*}
K_{a,b}=\frac{\kappa+1}{\eta}.
\end{equation*}
These findings explicitly relate the spectral properties of
the Hamiltonian to the two-point correlation functions.

\section{Entries of the correlation matrix for the nearest-neighbor case}
\label{nnEntries}

In this appendix, we make the evaluation of the entries of
$V^{1/2}$ and $V^{-1/2}$ of the important
case of nearest-neighbor interactions specific.
The power series expansion of the square root is given by
\begin{gather*}
    \left(1-x\right)^{\pm\frac{1}{2}}= 1\mp\sum_{k=1}^\infty a^\pm_k x^k,\quad
    a^-_k=\prod_{l=1}^k\frac{2l-1}{2l},\\
    a^-_k \ge \frac{a^-_k}{2k-1}=a^+_k \ge \frac{1}{2^k(2k-1)},
\end{gather*}
which is valid for $|x|<1$, i.e.,
the positivity constraint $0<c<1/2D$ allows us to write
\begin{eqnarray*}
(\lambda_{\vec{k}}(V))^{\pm \frac{1}{2}}&=&
1\mp\sum_{l=1}^\infty a^\pm_l\Big(2c\sum_{\delta=1}^D\cos\big(\frac{2\pi k_\delta}{n}\big)\Big)^l\\
&=&
1\mp\sum_{l=1}^\infty
a^\pm_lc^l
\Big(\sum_{\delta=1}^D\me^{\frac{2\pi\mi k_\delta}{n}}+\me^{-\frac{2\pi\mi k_\delta}{n}}\Big)^l.
\end{eqnarray*}
Using the multinomial theorem, we have
\begin{equation*}
\Big(\sum_{\delta=1}^D\me^{\frac{2\pi\mi k_\delta}{n}}+\me^{-\frac{2\pi\mi k_\delta}{n}}\Big)^l=
\sum l!\prod_{\delta=1}^D\frac{\big(\me^{\frac{2\pi\mi k_\delta}{n}}+\me^{-\frac{2\pi\mi k_\delta}{n}}\big)^{n_\delta}}{n_\delta!},
\end{equation*}
where the sum runs over all $n_\delta$ with $\sum_\delta n_\delta=l$.
Now, applying the binomial theorem
\begin{equation*}
\big(\me^{\frac{2\pi\mi k_\delta}{n}}+\me^{-\frac{2\pi\mi k_\delta}{n}}\big)^{n_\delta}=\sum_{r=o}^{n_\delta}
\binom{n_\delta}{r}\me^{\frac{2\pi\mi k_\delta(n_\delta-2r)}{n}}.
\end{equation*}
Substituting all the above into Eq.~(\ref{entries}), we find
\begin{equation*}
[V^{\pm 1/2}]_{\vec{i},\vec{j}}=
\delta_{\vec{i},\vec{j}}\mp\sum_{l=1}^\infty a_l^\pm c^l
\sum l!\prod_{\delta=1}^Df(d_\delta,n_\delta),
\end{equation*}
where $d_\delta=i_\delta-j_\delta$ and
\begin{equation*}
f(d_\delta,n_\delta)=
\sum_{r=0}^{n_\delta}\frac{1}{(n_\delta-r)!r!}\sum_{k=1}^n
\frac{\me^{2\pi\mi k(d_\delta+n_\delta-2r)/n}}{n}.
\end{equation*}
To the sum over $r$ only terms with $d_\delta+n_\delta-2r=zn$ for some $z\in\zz$ contribute.
We thus arrive at the following expression for the entries of $V^{\pm 1/2}$
\begin{equation*}
	[V^{\pm 1/2}]_{\vec{i},\vec{j}}=\\
	\delta_{\vec{i},\vec{j}}\mp\sum_{l=1}^\infty a_l^\pm c^l\!\!\!\!
	\sum_{\sum_\delta n_\delta=l}\!\!\!
	l!\prod_{\delta=1}^D\!\!\sum_{\substack{r=0\\
	d_\delta+n_\delta-2r=zn}}^{n_\delta}
	\!\!\!\!\!\!\!\!\!\!\!\!\!\frac{1}{(n_\delta-r)!r!},
\end{equation*}
i.e., for nearest-neighbor interactions the product 
$-[V^{-1/2}]_{\vec{i},\vec{j}}[V^{+1/2}]_{\vec{i},\vec{j}}$ is always positive for $\vec{i}\ne\vec{j}$. This does not hold in general
and makes it difficult to obtain explicit bounds on the entries of $V^{\pm 1/2}$ for more
general interactions $V$.
To obtain a lower bound we keep only the term $l=d(\vec{i},\vec{j})$
and $n_\delta=|d_\delta|$. The restriction on $r$ is then fulfilled for $r=0$ ($r=|d_\delta|$) if $d_\delta<0$ ($d_\delta>0$), yielding
\begin{eqnarray*}
|[V^{\pm 1/2}]_{\vec{i},\vec{j}}|&\ge&
a_{d(\vec{i},\vec{j})}^\pm c^{d(\vec{i},\vec{j})}
\frac{d(\vec{i},\vec{j})!}{\prod_{\delta=1}^D|i_\delta-j_\delta|!}\\
&\ge&
\left(\frac{c}{2}\right)^{d(\vec{i},\vec{j})}
\frac{d(\vec{i},\vec{j})!}{(2d(\vec{i},\vec{j})-1)\prod_{\delta=1}^D|i_\delta-j_\delta|!}.
\end{eqnarray*}
It is also possible to obtain an upper bound that is tighter than the one derived for
finite-ranged interactions: The elements of $V^{\pm 1/2}$ are symmetric under $\vec{i}-\vec{j}\rightarrow n\vec{e}_\delta-(\vec{i}-\vec{j})$, where $\vec{e}_\delta$ is a unit vector along dimension $\delta$. Thus, we can
demand $-n/2\le i_\delta-j_\delta\le n/2$. Then we find that $n_\delta$ has to
be larger or equal to $|i_\delta-j_\delta|$ otherwise the restriction on $r$ can not be fulfilled. This
in turn means that $l$ has to be larger or equal to $d(\vec{i},\vec{j})$.
We then obtain an upper bound by
summing all terms in the sum over $r$ regardless of the given restriction, yielding
\begin{eqnarray*}
    |[V^{\pm 1/2}]_{\vec{i},\vec{j}}| &\le& \sum_{l\ge
    d(\vec{i},\vec{j})} a_l^\pm c^l \sum_{\sum_\delta n_\delta=l}
    l!\prod_{\delta=1}^D \frac{2^{n_\delta}}{n_\delta !}\\
    &=& \sum_{l\ge d(\vec{i},\vec{j})} a_l^\pm (2cD)^l\\
    &=&\sum_{l=0}^\infty a_{l+d(\vec{i},\vec{j})}^\pm
    (2cD)^{l+d(\vec{i},\vec{j})}\\
    &\le& a_{d(\vec{i},\vec{j})}^\pm (2cD)^{d(\vec{i},\vec{j})} \sum_{l=0}^\infty
    (2cD)^l\\
    &=& \frac{a_{d(\vec{i},\vec{j})}^\pm}{1-2cD} (2cD)^{d(\vec{i},\vec{j})}.
\end{eqnarray*}
\section{Enumerating the relevant terms in the area law}
\label{abzaehlerei}
\begin{figure}[tb]
\begin{center}
\includegraphics[scale=\oscillatorscale]{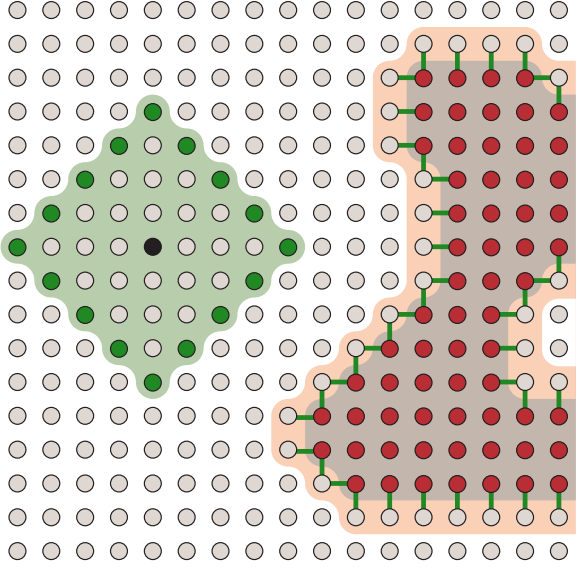}
\end{center}
\caption{Visualization of the enumeration of $N_{l}$ as in 
Appendix~\ref{abzaehlerei}. As before, oscillators belonging 
to the distinguished (shaded) region $I$ 
are marked~\usebox{\inneroscillatorbox}, the 
outside ones are shown as~\usebox{\outeroscillatorbox}. 
$\partial O$ is shown as the orange shaded area.
$M_{4}(\vec{i})=\{\vec{j}\in C|d(\vec{i},\vec{j})\leq 4\}$ 
for a certain oscillator~\usebox{\blackoscillatorbox} is shaded 
green, its surface oscillators $m_{4}(\vec{i})=\{\vec{j}\in C|d(\vec{i},\vec{j}) = 4\}$ 
are depicted by ~\usebox{\greenoscillatorbox}.}
\end{figure}
We start by identifying the set of oscillators that can contribute
to $N_l$, $l>1$,
\begin{equation*}
N_l=\sum_{\vec{j}\in O}\sum_{\substack{\vec{i}\in I\\ d(\vec{i},\vec{j})=l}}1.
\end{equation*}
Oscillators $\vec{j}\in O$ can only contribute if their distance to the boundary $\partial O$ is not larger than $l-1$,
\begin{equation*}
\partial O=\left\{\vec{j}\in O \;|\; \exists \vec{i}\in I: d(\vec{i},\vec{j})=1\right\}.
\end{equation*}
Thus, we can restrict the sum over $O$ to the set $A_l$
\begin{equation*}
A_l=\bigcup_{\vec{i}\in\partial O}\left\{\vec{j}\in O \;|\; d(\vec{i},\vec{j})\le l-1\right\},
\end{equation*}
i.e., we can write
\begin{equation*}
N_l=\sum_{\vec{o}\in A_l}\sum_{\substack{\vec{i}\in I\\ d(\vec{i},\vec{o})=l}}1
\le \sum_{\vec{o}\in A_l}\sum_{\substack{\vec{i}\in C\\ d(\vec{i},\vec{o})=l}}1
\le |A_l|m_l,
\end{equation*}
where $m_l$ is the number of surface oscillators of a ball with radius $l$ within the metric $d$, i.e., $m_l\le 2(2l+1)^{D-1}$ for $l\geq 1$. Using the fact that $|\partial O|\le N_1 = s(I)$, $|A_l|$ can now be bounded
from
above in the following way
\begin{equation*}
|A_l|\le |\partial O|M_{l-1}\le s(I)M_{l-1},
\end{equation*}
where $M_l$ is the volume of a ball with radius $l$ within the metric $d$, i.e., $M_{l}\le (2l+1)^{D}$. To summarize, we have
\begin{equation*}
N_l
\le
2(2l-1)^{D}(2l+1)^{D-1}s(I)\le 2(2l)^{2D-1}s(I).
\end{equation*}


\begin{thebibliography}{99}


\bibitem{Summers W 85}
    S.~J.~Summers and R.~F.~Werner, Phys.\ Lett.~A {\bf 110}, 257
    (1985).

\bibitem{Stelmachovic B 01} P.~Stelmachovic and V.~Buzek,
        Phys.\ Rev.~A {\bf 70}, 032313 (2004).

\bibitem{Osterloh}
        A.~Osterloh, L.~Amico, G.~Falci, and R.~Fazio,
        Nature {\bf 416}, 608 (2002).

\bibitem{NielsenPhase}
        T.~J.~Osborne and M.~A.~Nielsen,
        Phys.\ Rev.\ A {\bf 66}, 032110 (2002).

\bibitem{Audenaert EPW 02}
        K.~Audenaert, J.~Eisert, M.~B.~Plenio,
    and R.~F.~Werner, Phys.\
        Rev.\ A {\bf 66}, 042327 (2002).

\bibitem{Latorre1}
        G.~Vidal, J.~I.~Latorre, E.~Rico, and A.~Kitaev,
         Phys.\ Rev.\ Lett.\ {\bf 90}, 227902 (2003),
         \arxiv{quant-ph/0211074};
        J.~I.~Latorre, E.~Rico, and G.~Vidal,
        Quant.\ Inf.\ Comp.\ {\bf 4}, 048 (2004),
        \arxiv{quant-ph/0304098}.

\bibitem{Korepin1}
         B.-Q.~Jin and V.~E.~Korepin,
    J.\ Stat.\ Phys.\ {\bf 116}, 79 (2004);
        A.~R.~Its, B.-Q.~Jin, and V.~E.~Korepin,
        J.\ Phys.\ A {\bf  38},  2975 (2005).

\bibitem{Fannes}
	M.\ Fannes, B.\ Haegeman, and M.\ Mosonyi,
	 J.\ Math.\ Phys.\ {\bf 44}, 6005 (2003).
	 
\bibitem{AreaPRL}
        M.~B.~Plenio, J.~Eisert, J.~Drei\ss{}ig,
        and M.~Cramer, Phys.\ Rev.\ Lett.\ {\bf 94},
        060503 (2005),
        \arxiv{quant-ph/0405142}.

\bibitem{Eisert P 03}
    J.~Eisert and M.~B.~Plenio, Int.\ J.\ Quant.\ Inf.\ {\bf 1}, 479
    (2003).

\bibitem{Plenio V 98}
    M.~B.~Plenio and V.~Vedral, Contemp.\ Phys.\ {\bf 39}, 431 (1998).

\bibitem{Plenio V 05}
    M.~B.~Plenio and S.~Virmani,
    \arxiv{quant-ph/0504163}.

\bibitem{Keating}
        J.~P.~Keating and F.~Mezzadri,
    Phys.\ Rev.\ Lett.\ {\bf 94},
        050501 (2005);
        J.~P.~Keating and F.~Mezzadri,
        Commun.\ Math.\ Phys.\ {\bf  252}, 543
        (2004).

\bibitem{Cardy}
        P.~Calabrese and J.~Cardy,
        J.\ Stat.\ Mech.\ {\bf 06}, 002 (2004),
        \arxiv{hep-th/0405152}.

\bibitem{GraphStates}
        M.~Hein, J.~Eisert, and H.~J.~Briegel, Phys.\ Rev.\ A
        {\bf 69}, 062311 (2004).

\bibitem{Pachos P 04} 
	J.~K.~Pachos and M.~B.~Plenio, 
	Phys.~Rev.
~Lett. {\bf 93}, 056402 (2004).

\bibitem{Stabil}
        A.~Hamma, R.~Ionicioiu, and P.~Zanardi,
        Phys.\ Rev.\ A {\bf 71}, 022315 (2005).

\bibitem{DMRG}
    F.~Verstraete and J.~I.~ Cirac, \arxiv{cond-mat/0407066}.

\bibitem{Wolf VC 04}
    M.~M.~Wolf, F.~Verstraete, and J.~I.~Cirac,
    Phys.\ Rev.\ Lett.\ {\bf 92}, 087903 (2004).

\bibitem{Botero R 04}
    A.~Botero and B.~Reznik,
    Phys.\ Rev.\ A {\bf 70}, 052329  (2004);
    B.~Reznik, A.~Retzker, and J.~Silman,
    J.\ Mod.\ Opt.\ {\bf 51}, 833 (2004);
    B.~Reznik, Found.~Phys.~{\bf 33}, 167 (2003).

\bibitem{Peschel}
        I.~Peschel,
    J.~Stat.\ Mech.\ Th.\ E
    {\bf P12} (2004).

\bibitem{Beenakker}
        N.~Lambert,
        C.~Emary, and T.~Brandes, Phys.\ Rev.\ Lett.\
    {\bf 92}, 073602 (2004).

\bibitem{Longrange}
        W.~D{\"u}r, L.~Hartmann, M.~Hein, M.~Lewenstein,
        and H.~J.~Briegel, Phys.\ Rev.\ Lett.\
        {\bf 94}, 097203 (2005).

\bibitem{Wolf}
    M.~M.~Wolf, \arxiv{quant-ph/0503219}.

\bibitem{Klich}
    D.~Gioev and I.~Klich, \arxiv{quant-ph/0504151}.

\bibitem{Orus}
        R.~Orus, \arxiv{quant-ph/0501110}.

\bibitem{Fine}
        J.~I.~Latorre, C.~A.~L{\"u}tken, R.~Rico, and G.~Vidal,
        Phys.\ Rev.\ A {\bf 71}, 034301 (2005),
        \arxiv{quant-ph/0404120}.

\bibitem{Callan W 94}
    C.~Callan and F.~Wilczek, Phys.\ Lett.\ B {\bf 333},
    55 (1994),
    \arxiv{hep-th/9401072}.

\bibitem{Bekenstein 72}
    J.~D.~Bekenstein, Lett.\ Nuovo Cim.\ {\bf 4},
    737 (1972).

\bibitem{Bekenstein 73}
    J.~D.~Bekenstein, Phys.\ Rev.\ D {\bf 7}, 2333 (1973).

\bibitem{Bekenstein 04}
    J.~D.~Bekenstein, Contemp.\ Phys.\ {\bf 45},
    31 (2004).

\bibitem{Bombelli KLS 86}
    L.~Bombelli, R.~Koul, J.~Lee, and R.~Sorkin, Phys.\ Rev.\ D {\bf 34}, 373 (1986).

\bibitem{Srednicki 93}
    M.~Srednicki, Phys.\ Rev.\ Lett.\ {\bf 71}, 66
    (1993).

\bibitem{Holzhey LW 95}
    C.~Holzhey, F.~Larsen, and F.~Wilczek, Nucl.\
    Phys.\ B {\bf 424}, 443 (1995),
    \arxiv{hep-th/9403108}.

\bibitem{Preskill}
    T.~M.\ Fiola,  J.\ Preskill,  A.\ Strominger,
    and S.~P.\ Trivedi, Phys.\ Rev.\ D {\bf 50}, 3987 (1994).

\bibitem{CardyPeschel}
    J.~Cardy and I.~Peschel,
    Nucl.\ Phys.\ B {\bf 300}, 377 (1988).

\bibitem{Neg}
    K.~Zyczkowski, P.~Horodecki, A.~Sanpera, and M.~Lewenstein,
    Phys.\ Rev.\ A {\bf 58}, 883 (1998);
    J.~Eisert and M.~B.~Plenio, J.~Mod.\ Opt.\ {\bf 46}, 145 (1999);
    J.~Eisert (PhD thesis, Potsdam, February 2001);
    G.~Vidal and R.~F.~Werner, Phys.\ Rev.\ A {\bf 65}, 032314 (2002);
    K.~Audenaert, M.~B.~Plenio, and J.~Eisert,
    Phys.\ Rev.\ Lett.\ {\bf 90}, 027901 (2003); M.~B.~Plenio,
    \arxiv{quant-ph/0505071}.

\bibitem{InfContinuous}
    J.~Eisert, C.~Simon, and M.~B.~Plenio,
    J.\ Phys.\ A {\bf 35}, 3911  (2002).

\bibitem{Matrix}
    R.~A.~Horn and C.~R.~Johnson,
    {\it Matrix Analysis} (Cambridge University
    Press,
    Cambridge, 1985).

\bibitem{Hashing}
    I.~Devetak and A.~Winter,
    Proc.\ R.\ Soc.\ Lond.\ A {\bf  461},  207 (2005).

\bibitem{Benzi G 99}
    M.~Benzi and G.~H.~Golub,
    BIT Numerical Mathematics {\bf 39}, 417 (1999).

\bibitem{plusminus}
    Note that the above choice of $c>0$ does
    not restrict
    generality: clearly, there exists a diagonal matrix $M\in
    SO(n^D)$, the elements of which are $\pm1$, such that
    \begin{equation*}
        M V_{\delta+1} M^T \eqqcolon V'_{\delta+1}=
        \circulant(V'_\delta, c \id_{n^\delta},0,\dots,0, c\id_{n^\delta})
    \end{equation*}
    and $V_1'\coloneqq \circulant(1,c,0,\dots,0,c)$. Hence, there exists a
    local symplectic transformation $M\oplus M$ relating the
    Hamiltonian with $c<0$ to the one with $c>0$.

\bibitem{Wehrl}
    A.~Wehrl, Rev.~Mod.\ Phys.\ {\bf 50}, 221 (1978).

\bibitem{Callaway}
	D.J.E.\ Callaway,
	Phys.\ Rev.\ E {\bf 53} 3738 (1996).
\end{thebibliography}
\end{document}